\documentclass[12pt]{article}
\usepackage{amsfonts}
\usepackage{amsmath, amsthm}
\usepackage{epsfig}
\usepackage{algorithm}
\usepackage{algorithmic}
\usepackage{epic}
\usepackage[a4paper]{geometry}
\usepackage[colorlinks=true,urlcolor=blue,citecolor=red,linkcolor=blue,bookmarks=true]{hyperref}

\usepackage{amsmath,verbatim,color,amssymb,epsfig}
\usepackage{bm}
\usepackage{amsfonts}
\usepackage{epsfig}
\usepackage{changepage}
\usepackage{multirow}
\usepackage{graphicx}
\usepackage{array}
\usepackage[table]{xcolor}
\definecolor{Gray}{gray}{0.80}
\usepackage{lscape}
\usepackage[round,semicolon,authoryear]{natbib}
\usepackage[normalem]{ulem}
\usepackage{caption}
\begin{document}
\def\eqx"#1"{{\label{#1}}}
\def\eqn"#1"{{\ref{#1}}}

\makeatletter 
% make @ act like a letter
\@addtoreset{equation}{section}
\makeatother  
% make @ act like a non-letter

\def\yincomment#1{\vskip 2mm\boxit{\vskip 2mm{\color{red}\bf#1} {\color{blue}\bf --Yin\vskip 2mm}}\vskip 2mm}
\def\lincomment#1{\vskip 2mm\boxit{\vskip 2mm{\color{blue}\bf#1} {\color{black}\bf --Lin\vskip 2mm}}\vskip 2mm}
\def\squarebox#1{\hbox to #1{\hfill\vbox to #1{\vfill}}}
\def\boxit#1{\vbox{\hrule\hbox{\vrule\kern6pt
          \vbox{\kern6pt#1\kern6pt}\kern6pt\vrule}\hrule}}

\def\theequation{\thesection.\arabic{equation}}
\newcommand{\ds}{\displaystyle}

\newcommand{\bJ}{\mbox{\bf J}}
\newcommand{\bF}{\mbox{\bf F}}
\newcommand{\bM}{\mbox{\bf M}}
\newcommand{\bR}{\mbox{\bf R}}
\newcommand{\bZ}{\mbox{\bf Z}}
\newcommand{\bX}{\mbox{\bf X}}
\newcommand{\bx}{\mbox{\bf x}}
\newcommand{\bQ}{\mbox{\bf Q}}
\newcommand{\bH}{\mbox{\bf H}}
\newcommand{\bh}{\mbox{\bf h}}
\newcommand{\bz}{\mbox{\bf z}}
\newcommand{\ba}{\mbox{\bf a}}
\newcommand{\be}{\mbox{\bf e}}
\newcommand{\bG}{\mbox{\bf G}}
\newcommand{\bB}{\mbox{\bf B}}
\newcommand{\bb}{\mbox{\bf b}}
\newcommand{\bA}{\mbox{\bf A}}
\newcommand{\bC}{\mbox{\bf C}}
\newcommand{\bI}{\mbox{\bf I}}
\newcommand{\bD}{\mbox{\bf D}}
\newcommand{\bU}{\mbox{\bf U}}
\newcommand{\bc}{\mbox{\bf c}}
\newcommand{\bd}{\mbox{\bf d}}
\newcommand{\bs}{\mbox{\bf s}}
\newcommand{\bS}{\mbox{\bf S}}
\newcommand{\bV}{\mbox{\bf V}}
\newcommand{\bv}{\mbox{\bf v}}
\newcommand{\bW}{\mbox{\bf W}}
\newcommand{\bw}{\mbox{\bf w}}
\newcommand{\bg}{\mbox{\bf g}}
\newcommand{\bu}{\mbox{\bf u}}
\def\bb{{\bf b}}

\newcommand{\bcU}{\boldsymbol{\cal U}}
\newcommand{\bbeta}{\boldsymbol{\beta}}
\newcommand{\bdelta}{\boldsymbol{\lambda}}
\newcommand{\bDelta}{\boldsymbol{\lambda}}
\newcommand{\boldeta}{\boldsymbol{\eta}}
\newcommand{\bxi}{\boldsymbol{\xi}}
\newcommand{\bGamma}{\boldsymbol{\Gamma}}
\newcommand{\bSigma}{\boldsymbol{\Sigma}}
\newcommand{\balpha}{\boldsymbol{\alpha}}
\newcommand{\bOmega}{\boldsymbol{\Omega}}
\newcommand{\btheta}{\boldsymbol{\theta}}
\newcommand{\bmu}{\boldsymbol{\mu}}
\newcommand{\bnu}{\boldsymbol{\nu}}
\newcommand{\bgamma}{\boldsymbol{\gamma}}

\newtheorem{thm}{Theorem}
%[section]
\newtheorem{lem}{Lemma}[section]
\newtheorem{rem}{Remark}[section]
\newtheorem{cor}{Corollary}[section]
\newcolumntype{L}[1]{>{\raggedright\let\newline\\\arraybackslash\hspace{0pt}}m{#1}}
\newcolumntype{C}[1]{>{\centering\let\newline\\\arraybackslash\hspace{0pt}}m{#1}}
\newcolumntype{R}[1]{>{\raggedleft\let\newline\\\arraybackslash\hspace{0pt}}m{#1}}

\newcommand{\tabincell}[2]{\begin{tabular}{@{}#1@{}}#2\end{tabular}}

\baselineskip=24pt
\begin{center}
 {\Large \bf Time-to-Event Model-Assisted Designs to Accelerate Phase I Clinical Trials}
\end{center}

%\vspace{2mm}
\begin{center}
{\bf Ruitao Lin and Ying Yuan }
\end{center}

\begin{center}

Department of Biostatistics, The University of Texas MD Anderson Cancer Center,\\
 Houston, TX 77030, USA.

{\em email}: ruitaolin@gmail.com ~~~ yyuan@mdanderson.org \\

\bigskip
\noindent{Abstract}
\end{center}
Two useful strategies to speed up drug development are to increase the patient accrual rate and use novel adaptive designs. Unfortunately, these two strategies often conflict when the evaluation of the outcome cannot keep pace with the patient accrual rate and thus the interim data cannot be observed in time to make  adaptive decisions. A similar logistic difficulty arises when the outcome is of late onset. Based on a novel formulation and approximation of the likelihood of the observed data, we propose a general methodology for model-assisted designs to handle toxicity data that are pending due to fast accrual or late-onset toxicity,  and facilitate seamless decision making in phase I dose-finding trials. The dose escalation/de-escalation rules of the proposed time-to-event model-assisted designs can be tabulated before the trial begins, which greatly simplifies trial conduct in practice compared to that under existing methods. We show that the proposed designs have desirable finite and large-sample properties and yield performance that is superior to that of more complicated model-based designs. We provide user-friendly software for implementing the designs.

\vspace{0.5cm}

\newpage
\section{Introduction}
Drug development enterprises are struggling because of unsustainably long development cycle and high costs. The pharmaceutical industry and regulatory agencies both recognize the urgency and necessity of speeding up drug development. Toward that goal, two common strategies are to increase the accrual rate to shorten the trial duration \citep{Dilts08} and to use novel adaptive designs for more efficient decision making \citep{kairalla2012}. These two strategies, unfortunately, often conflict. The majority of adaptive designs require that the endpoint is quickly ascertainable, such that by the time interim decisions to be made, the outcomes of patients already enrolled in the trial have been fully ascertained. When the accrual rate is fast, some enrolled patients may not have completed their outcome assessment by the interim decision making time, which causes a major logistic difficulty for implementing adaptive trial designs. This is particularly true for phase I trials, where adaptive decisions of dose escalation and de-escalation are mandated after each patient or patient cohort is treated. For example, in a trial where the dose limiting toxicity (DLT) takes up to 28 days to evaluate and patients are treated in cohorts of 3 patients, if the accrual rate is 2 patients per week, an average of 5 new patients will be accrued while the investigators wait to evaluate the outcomes of the 3 previously enrolled patients. The question is then: How can new patients receive timely treatment when the previous patients' outcomes are pending? 

The same difficulty arises when the DLT is of late onset and requires a long assessment window to be ascertained. For example, if the DLT assessment window of a new agent is 3 months, given the accrual rate of 3 patients per month, then an average of 6 new patients will be accrued while investigators wait to evaluate the DLT outcomes of the 3 previously enrolled patients. And still the investigators must determine how to provide the new patients with timely treatment. The problem of late-onset toxicity is particularly common and important in the era of targeted therapy and immunotherapy. A recent study reported that in 36 clinical trials of molecularly targeted agents, 
more than half of the grade 3 or 4 toxicity events occurred after the first treatment cycle \citep{postel2011}. Immunotoxicity is often of late onset,  for instance, endocrinopathies have been observed between post-treatment weeks 12 and 24 \citep{weber2015, june2017}. 

The fundamental issue associated with fast accrual and late-onset toxicity is that at the interim decision time, some patients' DLT data are pending (i.e., unknown), which causes difficulty in making adaptive decisions. 
 A number of novel phase I trial designs have been proposed to accommodate pending data such that adaptive decisions can be made in real time.  \cite{cheung2000} proposed the time-to-event continual reassessment method (TITE-CRM), where the likelihood of each patient is weighted by his/her follow-up proportion. This weighted likelihood approach has been applied to other designs such as the escalation with overdose control method \citep{babb1998,mauguen2011}. 
Taking a different perspective,   \cite{yuan2011}, \cite{liu2013} and \cite{jin2014} treated the pending DLT data as a missing data problem, and proposed the expectation--maximization algorithm and Bayesian data augmentation method to facilitate real-time decision making. \cite{lin2017a} used the Kaplan--Meier estimator to impute the missing toxicity data with the fractional DLT information. These model-based designs yield excellent operating characteristics, but their use in practice has been limited because they are often perceived by practitioners as difficult to understand, due to the blackbox-style of decision making, and complicated to implement, because of the requirement of repeated model fitting and estimation. Thus, in practice, the algorithm-based rolling six (R6) design  \citep{skolnik2008} is often used even though its performance is inferior to that of the model-based designs \citep{zhao2011}. 
To implement R6,  investigators count the number of patients with DLTs, the number of patients without DLTs and the number of patients with pending outcomes, and then use the prespecified decision table to determine the dose  for the next new cohort, in a fashion similar to the 3+3 design.

The goal of this article is to develop new model-assisted phase I designs that can adaptively handle 
pending DLT data due to fast accrual and/or late-onset toxicity. 
The proposed designs are transparent and can be implemented in a simple way like the algorithm-based design (e.g., the R6 design), but they have desirable statistical properties and yield performance that is comparable to that of the model-based designs (e.g., TITE-CRM). 
The proposed designs allow users to tabulate the dose escalation and de-escalation rules before the trial begins. To conduct the trial, there is no complicated model fitting and calculation, investigators only need to look up the decision table to make the dose-assignment decisions. Simulation studies show that albeit simplistic, the proposed design yields excellent operating characteristics compared to those of the more complicated model-based design.

Our approach is built upon the framework of model-assisted designs \citep{yan2017, zhou2018a}, a novel class of designs that 
combine the simplicity of algorithm-based designs with the superior performance of model-based designs. Model-assisted designs 
use a probability model for efficient decision making like model-based designs, while their dose escalation and de-escalation rules  can be pre-tabulated before the onset of a trial, as with algorithm-based designs.
\cite{ivanova2007} developed the cumulative cohort design based on the asymptotic distribution
of patient allocation of the up-and-down design using the Markov chain theory.  
 \cite{ji2010} proposed the modified toxicity probability interval (mTPI) design that utilizes unit probability mass (UPM) to guide the dose assignment. 
\cite{liu2015} developed the Bayesian optimal interval (BOIN) design that makes the decision of dose escalation and de-escalation by simply comparing
the observed toxicity rate at the current dose with two prespecified  boundaries that are optimized to minimize the incorrect decision of dose assignment.
\cite{yan2017} noted the overdosing issue of the mTPI design due to the use of UPM and  proposed the keyboard design as a seamless improvement of the mTPI to achieve higher accuracy of identification of the maximum tolerated dose (MTD) and better overdose control.  The model-assisted designs have been extended to drug-combination trials \citep{zhang2016, lin2016, pan2018}, to account for toxicity grades \citep{mu2018}, and for phase I-II trials \citep{lin2017}. For overviews and comparison of model-assisted designs, see \cite{zhou2018a} and \cite{zhou2018b}. Because of their simplicity and good performance, model-assisted designs have been increasingly used in practice. However, these designs cannot handle the issue of fast accrual or late-onset toxicity, and they all require that accrued patients have completed DLT assessment before treating the next patients.

Our study is motivated by a phase I trial planned at MD Anderson Cancer Center. The objective is to find the MTD of a MEK inhibitor combined with 200mg pembrolizumab for treating patients with advanced melanoma. Four doses,  100mg, 125mg, 150mg, and 175mg, of the MEK inhibitor will be studied, administered orally twice daily on a schedule of 3 days on, 4 days off. The maximum sample size is 21 patients, treated in cohort sizes of 3. As the DLT of the treatment is expected to be of late onset, the clinical investigator set the DLT assessment window at 3 months. The DLT will be scored using the NCI common Terminology Criteria for Adverse Events, version 4.0. The accrual rate is expected to be 2 patients per month.

The remainder of this article is organized as follows. In Section  \ref{Sec:Method},
we formulate a new likelihood-based approach to account for both observed and pending DLT data and develop the TITE-keyboard design and study its theoretical properties. We also briefly discuss the development of other model-assisted designs.
As an illustration,
we apply the TITE-keyboard design to a phase I dose-finding
trial  in Section \ref{Sec:Trial}.
In Section \ref{Sec:Simu}, we examine the performance of
the new design based on simulation studies and make extensive comparisons
with existing methods.
Section \ref{Sec:Con} concludes with some remarks.
The Supplementary Material contains the proofs of theorems.

\section{Methodology}
\label{Sec:Method}

Consider a phase I dose-finding trial with $J$ prespecified doses and maximum sample size $N$. Let $p_j$ denote the toxicity probability of dose level $j$, $j=1,\ldots,J$, with $p_1<\cdots<p_J$, and $\phi$ denote the target DLT rate. The objective is to find the MTD, defined as the dose that has the DLT probability closest to $\phi$.  Patients are sequentially enrolled, and each patient is followed for a fixed period of time, say $\tau$, to assess the binary DLT outcome $x_i$. If DLT  is observed within the assessment window $(0,\tau)$, $x_i=1$; otherwise, $x_i=0$. Let $t_i$ denote the time to DLT for patients with $x_i=1$, where $0 \le t_i \le  \tau$.  The DLT assessment window $\tau$ is prespecified by clinicians such that it is expected to capture all DLTs relevant for the MTD determination. For many chemotherapies, $\tau$ is often taken as the first cycle of 28 days; whereas for agents expected to induce late-onset toxicity, e.g., some targeted or immunotherapy agents, $\tau$ can be several (e.g., 3 to 6) months or longer. 

\subsection{Keyboard design}
The proposed methodology to deal with fast accrual and late-onset toxicity is general and applies to all model-assisted designs, including keyboard, mTPI as well as BOIN. For ease of exposition, we illustrate our approach using the keyboard design, reviewed briefly as follows. The keyboard design starts by specifying a proper dosing interval ${\cal I}^*=(\phi-\delta_1, \phi+\delta_2)$, referred to as the ``target key",  which represents the range of toxicity probabilities that are close enough to the target $\phi$ so that they are regarded as acceptable in practice, where $\delta_1$ and $\delta_2$ are small constants, such as $\delta_1=\delta_2=0.05$. The keyboard design populates the interval toward both sides of the target key, forming a series of keys of equal width that span the range of 0 to 1. For example, given the target key of (0.25, 0.35), on its left side, we form 2 keys of width 0.1, i.e., (0.15, 0.25) and (0.05, 0.15); and on its right side, we form 6 keys of width 0.1, i.e., $(0.35, 0.45), \ldots, (0.85, 0.95)$. We denote the resulting intervals/keys as ${\cal I}_1, \ldots, {\cal I}_K$, and assume that the $k^*$th interval is the target key, i.e., ${\cal I}_{k^*} = {\cal I}^*$.

Suppose that at a particular point during the trial, $n_j$ patients have been treated at dose level $j$, and among them $y_j = \sum_{i=1}^{n_j} x_i $ patients experienced DLT.  The keyboard design assumes a beta-binomial model
\begin{eqnarray}
y_j \,|\, n_j, p_j & \sim & \mathrm{Binom}(n_j,p_j), \label{betabinom} \\
p_j & \sim & \mathrm{Beta}(1,1) \equiv \mathrm{Unif}(0,1). \notag
\end{eqnarray}
Given data $D_j = (n_j,  y_j)$ observed at dose level $j$,  the posterior distribution arises as
\begin{equation}
p_j \,|\, D_j \sim \mathrm{Beta}(y_j+1, n_j-y_j+1), \; \mathrm{for} \; j=1,\ldots,J. \label{bbpost}
\end{equation}
In contrast to model-based designs (e.g., the CRM), which model toxicity  across doses using a dose--toxicity curve model (e.g., a power or logistic model),  the keyboard design models toxicity at each dose independently, which simplifies the design and renders it possible to pre-tabulate the decision rules of dose escalation and de-escalation.  Modeling toxicity at each dose independently is an essential feature of the model-assisted designs: mTPI assumes the same beta-binomial model as above and BOIN only assumes the binomial model for $y_j$.

To make the decisions of dose escalation and de-escalation, where $j$ is the current dose level,  the keyboard design identifies the interval  that has the largest posterior probability (referred to as the ``strongest" key),  i.e., 
$$k_{s}  = \underset{k=1, \cdots, K}{\mathrm{arg~max}} \{ {\rm Pr}( p_j \in {\cal I}_k \,|\, D_j) \},$$
which can easily be evaluated based on the posterior distribution of $p_j$ given by equation (\ref{bbpost}). 
The keyboard design determines the next dose as follows:
If $k_{s}<k^*$, escalate the dose to level $j+1$;  if $k_{s}=k^*$, retain the current dose level $j$; if $k_{s}>k^*$, de-escalate the dose to level $j-1$.
This dose escalation/de-escalation process continues until the prespecified sample size $N$ is exhausted, and the MTD is selected as the dose for which the isotonic estimate  \citep{barlow1972} of the toxicity rate is closest to the target $\phi$.

The most appealing feature of the keyboard design is that its decision rule can be tabulated before the trial begins, which greatly simplifies the practical implementation of the design.  This is possible because given the maximum sample size $N$, the possible outcome $D_j=(n_j, y_j)$ is finite for $n_j=1, \cdots, N$ and $y_j=0, \cdots, n_j$, and given each of the possible outcomes, the strongest key and thus the dose escalation/de-escalation rule can be easily determined based on (\ref{bbpost}). This is also the reason why the decision rules of the mTPI can be enumerated. We note that the mTPI2 \citep{Guoji17}, based on the  concept of Ockham's razor, ends up being the same design as the keyboard design; thus, our approach described below is applicable to mTPI2 as well.

\subsection{Likelihood with pending DLT data}
When the accrual is fast or when there is late onset of DLT, the fundamental difficulty that cripples the keyboard and other model-assisted designs is that by the time of decision making,  say time $\kappa$, the $x_i$'s may not be observed for patients who have not completed their DLT assessment. The data actually observed are indicator variables $\tilde{x}_i$, $i=1, \cdots, n_j$, which indicate that the patient has experienced DLT ($\tilde{x}_i=1$) or not yet ($\tilde{x}_i=0$) by time $\kappa$. Clearly, $\tilde{x}_i=1$ implies ${x}_i=1$, but when $\tilde{x}_i=0$, ${x}_i$ can equal 0 or 1.

Let $\delta_i$ indicate that the toxicity outcome $x_i$  has been ascertained (i.e., $\delta_i=1$) or is still pending (i.e., $\delta_i=0$) by the decision time $\kappa$, and $u_i$ $(u_i\leq \tau$) denote the actual follow-up time for patient $i$ up to that moment. For a patient with $\delta_i=1$,  we have $\tilde{x}_i = x_i$, thus the likelihood is given by
\begin{equation}
 \Pr (\tilde{x}_i = x_i | \delta_i=1)=p_j^{x_i}(1-p_j)^{1-x_i}. \label{complike}
 \end{equation}  
For a patient with $\delta_i=0$, $x_i$ has not been ascertained yet and his/her DLT outcome is pending. These patients with pending outcome data are a mixture of two subgroups: patients who will not experience DLT (i.e., $x_i=0$), and patients who will experience DLT (i.e., $x_i=1$) but have not experienced it yet by the interim decision time (i.e., $u_i<t_i$). Note that $\tilde{x}_i$ only takes a value of 0 because once $\tilde{x}_i=1$ (i.e., the patient experiences DLT), $x_i$  becomes observable and $\delta_i=1$. Therefore, for a patient with pending data with $\delta_i=0$, the likelihood is given by
\begin{eqnarray*}
\Pr(\tilde{x}_{i}=0\mid \delta_i=0 ) & = & \Pr(x_{i}=0) \Pr(\tilde{x}_i=0 | x_{i}=0) +\Pr(x_{i}=1)\Pr(\tilde{x}_i=0 \mid x_{i}=1)\\
& = & \Pr(x_{i}=0)+\Pr(x_{i}=1)\Pr(t_{i}> u_{i}\mid x_{i}=1)\\
& = & 1-p_j + p_j \{1-\Pr(t_{i} \leq  u_{i}\mid x_{i}=1)\}\\
 & = & 1-p_jw_i,
\end{eqnarray*}
where $w_i=\Pr( t_{i}\leq u_{i} \mid x_{i}=1)$ can be interpreted as a weight, adjusting for the fact that the DLT outcome has not been ascertained yet. We discuss how to specify $w_i$ later.

Given the observed interim data ${D}^o_j = (\tilde{x}_1, \cdots, \tilde{x}_{n_j}, \delta_1, \cdots, \delta_{n_j})$  at the current dose level $j$,  the joint likelihood function  is given by 
\begin{eqnarray}\label{likelihood}
L(p_j\mid {D}_j^o) & \propto & \prod_{i=1}^{n_j}p_j^{\delta_{i}x_{i}}(1-p_j)^{\delta_{i}(1-x_{i})}(1-w_{i}p_j)^{1-\delta_{i}} \nonumber \\ 
 & = & p_j^{\tilde{y}_j}(1-p_j)^{m_j}\prod_{i=1}^{n_j}(1-w_{i}p_j)^{1-\delta_{i}}, \label{likelihood}
\end{eqnarray}
where $\tilde{y}_j=\sum_{i=1}^{n_j}\delta_ix_i$ is the number of patients who experienced DLT by the interim time $\kappa$, and $m_j=\sum_{i=1}^{n_j}\delta_i(1-x_i)$ is the number of patients who have completed the assessment without experiencing DLT. Despite some similarity (e.g., using weights), our likelihood function (\ref{likelihood}) differs from the one utilized in the TITE-CRM \citep{cheung2000}. In the TITE-CRM, as long as the patient's follow-up time $u_i< \tau$, that patient's likelihood will be weighted for a partial credit, no matter whether the patient has experienced DLT or not at that moment. This is odd, however, because once a patient has experienced DLT, his/her toxicity outcome $x_i$ has been ascertained. That patient's data should receive full credit and the standard binomial likelihood (\ref{complike}) should be used, even though the patient has not gone through the whole assessment window yet. In contrast, our approach only weights the observations from the patients whose $x_i$ is yet unknown, and for the patients whose $x_i$ has been ascertained, it uses the standard binomial likelihood.

Let $\pi(p_j)$ denote the prior distribution for $p_j$, e.g., $\pi(p_j) =  {\rm Beta}(1, 1)$. The posterior distribution $f(p_j\mid D_j^o)$ is given by 
\begin{equation*}
f(p_{j}\mid D_{j}^o)\propto \pi(p_j) L(p_j\mid {D}_j^o). 
\end{equation*}
Although this posterior distribution can be routinely sampled by the Markov chain Monte Carlo method and the strongest key thus can be  identified to determine dose assignment, the issue is that now it is impossible to enumerate the dose escalation and de-escalation rules before the trial begins. This is because the posterior depends on individual-level continuous variable $w_i$ and there is an infinite number of possible outcomes, prohibiting the enumeration of the decision rule. As a result, the keyboard design with pending DLT data loses its most appealing advantage. This issue also occurs for the other model-assisted designs.

\begin{figure}
\begin{center}
\includegraphics[width=2.4in,height=2.4in]{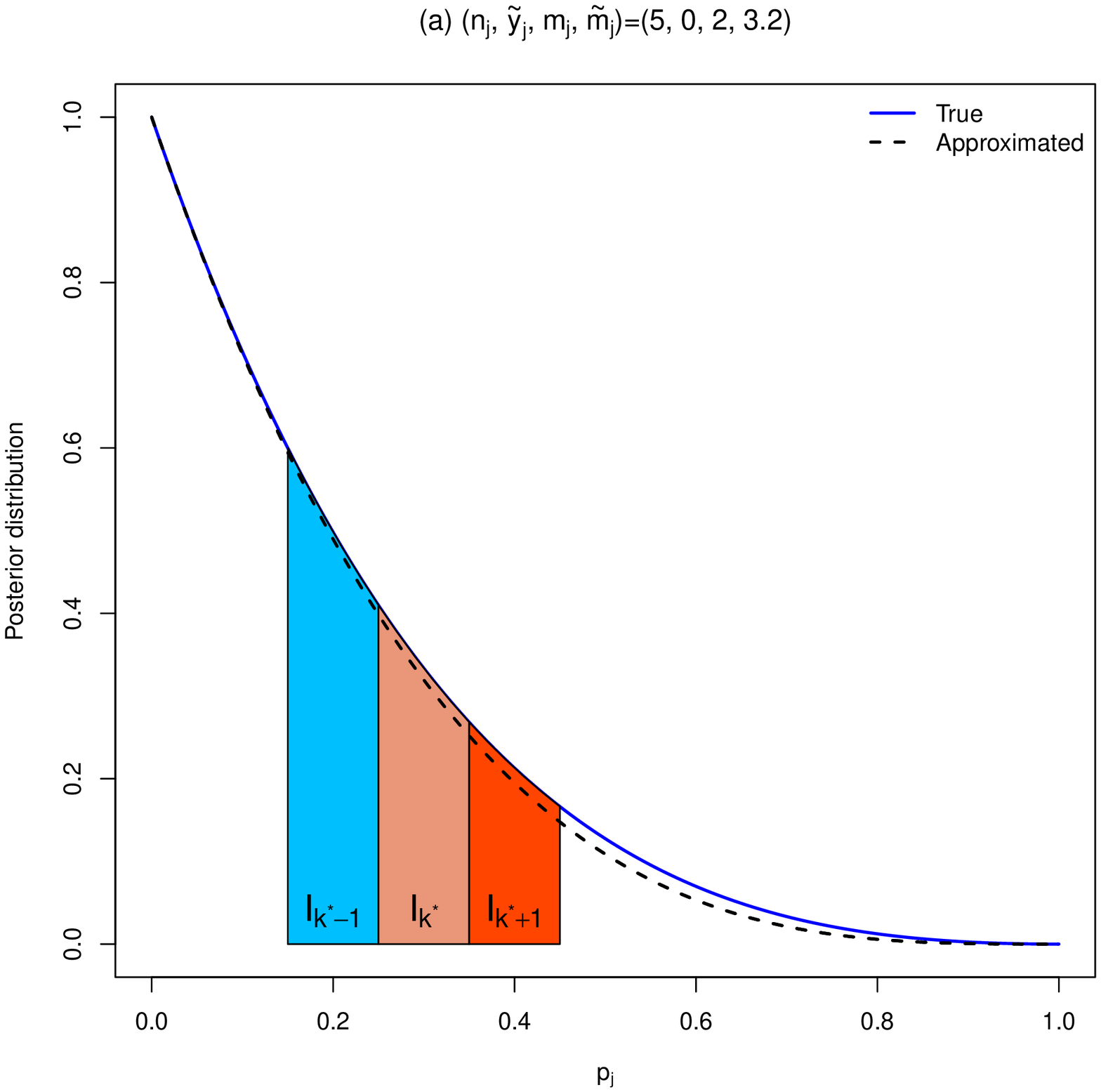}
\includegraphics[width=2.4in,height=2.4in]{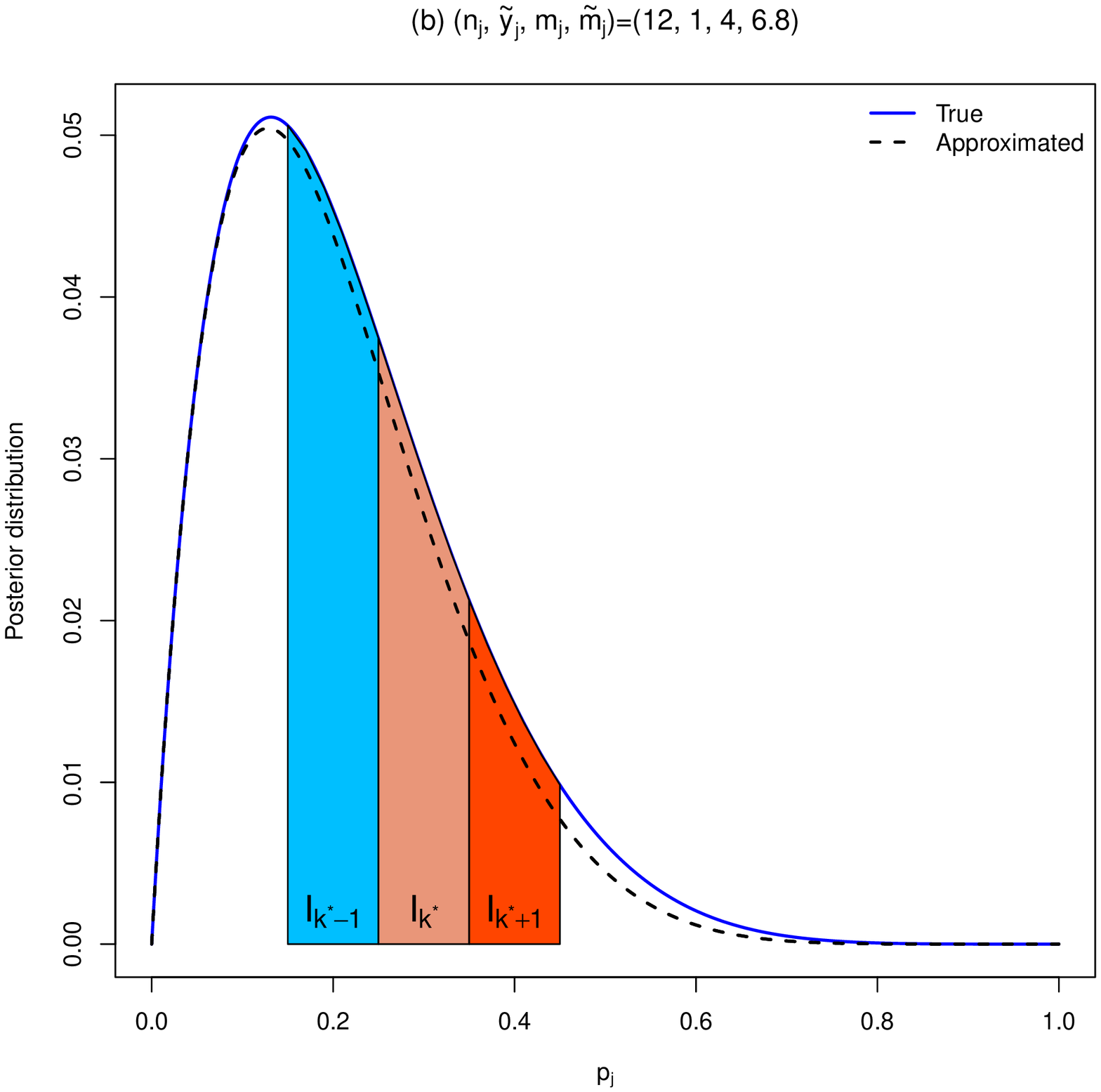}

\end{center}
\caption{The true and approximated posterior functions based on the observed data: (a) $n_j=5$ patients have been treated and only $m_j=2$ patients have finished the assessment without any DLT, and the weights for the remaining 3 patients are 0.3, 0.4, and 0.5; (b) $n_j=12$ patients have been treated, $\tilde{y}_j=1$ DLT has been observed,  $m_j=4$ patients have finished the assessment without any DLT, and the weights for the remaining 7 patients are $0.1,0.2,\ldots,0.7$. The prior distribution of $p_j$ is $p_j\sim{\rm Unif}(0,1)$; $\tilde{y}_j$ and $\tilde{m}_j$ respectively represent  the ``effective'' numbers of patients with DLT and patients without DLT by the interim time $\gamma$; $\mathcal{I}_{k^*}$ represents the target key. }
\label{approximation}
\end{figure}

To circumvent the aforementioned issue and maintain the simplicity of the keyboard design, we 
approximate the last term in  (\ref{likelihood}) as follows, 
\begin{equation}
 (1-w_{i}p_{j})^{1-\delta_{i}} \approx (1-p_{j})^{w_i(1-\delta_{i})}. \label{approx}
 \end{equation}
When $\delta_i=1$, $(1-w_{i}p_{j})^{0} =(1-p_{j})^{0}=1$; when $\delta_i=0$, 
we perform Taylor expansion of $(1-p_{j})^{w_i}$ at $p_j=0$,  resulting in
$$(1-p_{j})^{w_i}=1-w_ip_j+w_i(w_i-1)p_j^2+\cdots$$ 
Ignoring the second-order and higher terms, we obtain the  approximation (\ref{approx}). Thus, the likelihood (\ref{likelihood}) is approximated as
\begin{equation}
{L}(p_j\mid {D}^o_j) \propto p_{j}^{\tilde{y}_{j}}(1-p_{j})^{\tilde{m}_{j}},  \label{approxlik}
%\tilde{f}(p_{j}\mid D_{j})\propto p_{j}^{y_{j}}(1-p_{j})^{m_{j}+{\rm 
%TFP}_{j}},
\end{equation}
where 
\begin{equation}
\tilde{m}_{j}=m_j+\sum_{i=1}^{n_{j}}(1-\delta_{i})w_{i}.  \label{mj}
\end{equation}
This approximation is simple but powerful. It converts the non-regular likelihood  (\ref{likelihood}) into a binomial likelihood arising from  ``effective" binomial data $\tilde{D}_j = (\tilde{n}_j, \tilde{y}_j)$, where $\tilde{n}_j= \tilde{y}_j+ \tilde{m}_{j}$ is the ``effective" sample size, and $\tilde{m}_{j}$ is the ``effective" number of patients who have not experienced DLT.  As all model-assisted designs are based on the binomial likelihood, this means that these designs can be seamlessly extended to accommodate pending DLT data using the approximated likelihood (\ref{approxlik}), as demonstrated in Section \ref{tite-keyboard}. In addition, because  the approximated likelihood (\ref{approxlik}) depends on the aggregated value of the $w_i$'s, rather than the individual value of $w_i$, the approximation renders it possible to enumerate the decision rules for the resulting design, maintaining the most important feature of the model-assisted designs. Lastly, the following theorem indicates that (\ref{approxlik}) provides a very accurate approximation of the true likelihood (\ref{likelihood}). 
\begin{thm}\label{thm1}
Let $l(w_{i},\delta_i,p_{j})=(1-w_{i}p_{j})^{1-\delta_{i}}$ and  $\,\, \tilde{l}(w_{i},\delta_i,p_{j})=(1-p_{j})^{w_i(1-\delta_{i})}$. The approximation error is bounded by  \[
{d(w_{i},\delta_i,p_{j})=|l(w_{i},\delta_i,p_{j})-\tilde{l}(w_{i},\delta_i,p_{j})|}\leq(1-\beta_{j} p_{j})-(1-p_{j})^{\beta_{j}},
\]
where $\beta_j=\log\{-p_{j}/\log(1-p_{j})\}/\log(1-p_{j})$. Specifically, for any $p_j\leq0.4$, the approximation error $d(w_{i},\delta_i,p_{j})<0.0255$.
\end{thm}
The proof of Theorem \ref{thm1} is provided in the Supplementary Material. To illustrate the accuracy of the approximation, we consider two examples: (a) $n_j=5$ patients have been treated and only $m_j=2$ patients have finished the assessment without any DLT, and the weights $w_i$ for the remaining 3 patients are 0.3, 0.4, and 0.5, respectively; (b) $n_j=12$ patients have been treated, $y_j=1$ DLT has been observed,  $m_j=4$ patients have finished the assessment without any DLT, and the weights $w_i$ for the remaining 7 patients are $0.1,0.2,\ldots,0.7$, respectively. Given the prior $p_j\sim {\rm Unif}(0,1)$, Figure \ref{approximation} shows the true and approximated posterior distribution functions, indicating that the approximation is very accurate.

\subsection{Specifying the weight}
We consider three different schemes for specifying the weight $w_i$ that appears in the approximated likelihood (\ref{approxlik}).

{\bf (1) Uniform weight}\\
Following the TITE-CRM, the simplest way is to assume that the time-to-toxicity outcome is uniformly distributed over the assessment period $(0,\tau)$, i.e., $t_i\mid x_i=1 \sim {\rm Unif}(0,\tau)$, 
leading to 
\begin{equation}\label{unif}
w_i^{\rm u}=\Pr(t_{i}<u_{i}\mid x_{i}=1)=u_i/\tau.
\end{equation}
As a result, $w_i^{\rm u}$ can be interpreted as the follow-up proportion that patient $i$ has finished. As shown later in simulation, although the uniform scheme seems very restrictive, it yields remarkably robust performance. This was also observed by \cite{cheung2000} in the TITE-CRM, and based on that they recommended the uniform weight as the default for general use.

{\bf (2) Piecewise uniform weight}\\
A more flexible weighting scheme is to assume that the time to toxicity follows a piecewise uniform distribution, which partitions $(0, \tau)$ into several  intervals and assumes a uniform distribution within each interval. By increasing the number of partitions,  the piecewise uniform distribution can approximate any shape of the time-to-toxicity distribution. This makes it particularly useful to incorporate prior information on the time to toxicity, $t_i$.   For ease of exposition, we consider three partitions that are often adequate for practical use. We partition the assessment window $(0,\tau)$ into the initial part $(0,\tau/3)$, the middle part  $(\tau/3,2\tau/3)$ and the final part $(2\tau/3,\tau)$, and assume that $t_i$  is uniformly distributed in each interval. Let $(\upsilon_1,\upsilon_2,\upsilon_3)$ be the prior probability that the DLT would occur at the three parts of the assessment window, where $\upsilon_1+\upsilon_2+\upsilon_3=1$. 
For example, prior data may suggest that the DLT is more likely to occur late in the assessment window, in which case we can choose $\upsilon_3>\upsilon_2>\upsilon_1$. It then follows that
\begin{eqnarray*}
w_i^{\rm p}=\Pr(t_{i}<u_{i}\mid x_{i}=1) & = & \begin{cases}
3\upsilon_{1}u_{i}/\tau, & u_{i}\in(0,\tau/3),\\
\upsilon_{1}-\upsilon_{2}+3\upsilon_{2}u_{i}/\tau, & u_{i}\in(\tau/3,2\tau/3)\\
\upsilon_{1}+\upsilon_{2}-2\upsilon_{3}+3\upsilon_{3}u_{i}/\tau, & u_{i}\in(2\tau/3,\tau),
\end{cases},
\end{eqnarray*}
where $w_i^{\rm p}$ can be interpreted as the weighted follow-up probability that patient $i$ has completed the assessment time.   

{\bf (3) Adaptive weight}\\
The first two schemes determine $w_i$ based on prespecified prior distributions. The adaptive scheme estimates $w_i$ adaptively during the trial based on the accumulating interim data. Note that $t_i$ is the time to DLT for patients who have experienced DLT (i.e., $x_i=1$), thus its support is $(0, \tau)$. Hence, we assume that $t_i$ follows a scaled Beta distribution
\begin{eqnarray*}
t_{i} \mid x_i=1,\lambda,\gamma & \sim & \tau \times {\rm Beta}(\lambda,\gamma);\\
\lambda,\gamma & \sim & \pi(\lambda, \gamma),
\end{eqnarray*}
where $\pi(\lambda, \gamma)$ is the prior distribution of the unknown parameters $\lambda$ and $\gamma$, e.g., independent gamma prior, ${\rm Gamma}( 0.1, 0.1 )$.  As data are very sparse in phase I trials, to borrow information, we consider that given $x_i=1$, $t_i$ follows the same distribution across doses. Letting $D=\cup_{j=1}^J D_j$, the posterior estimate of $w_i$ is given by 
$$w_{i}^{\rm a}=\int_{0}^{\infty}\int_{0}^{\infty}\int_{0}^{u_{i} }g(t_{i}\mid\lambda,\gamma)f(\lambda,\gamma\mid D){\rm d}t_{i}{\rm d}\lambda{\rm d}\gamma,$$
where $f(\lambda,\gamma\mid D)$ is the posterior distribution of $\lambda$ and $\gamma$, and $g(t_{i}\mid\lambda,\gamma)$ is the density of $t_i$ based on the scaled Beta distribution. 
Because the adaptive schemes allows $w_i$ to adaptively change with the observed interim data, it is more flexible and theoretically more favorable. However, the numerical study (in Section 4)shows that the adaptive scheme provides minimal improvement over the first two schemes, but is substantially more complicated, largely because the data are too sparse to provide reliable estimates of the time to DLT.  Thus, we recommend the first two weighting schemes for general use.

\subsection{TITE-keyboard design \label{tite-keyboard}}
Application of the proposed methodology to the keyboard design is straightforward. We refer to the resulting design as the TITE-keyboard design. The decision rule of the TITE-keyboard design is almost the same as that of the keyboard design. The only change is that to make decisions of dose escalation and de-escalation, we  replace the complete data $D_j=(n_j, y_j)$, which are not observable when some DLT data are pending, with the ``effective" binomial data $\tilde{D}_j=(\tilde{n}_j, \tilde{y}_j)$ for calculating the posterior distribution of $p_j$ and identifying the strongest key. Once the strongest key is identified, the same dose escalation/de-escalation rule is used to guide the dose transition.  

Compared to the model-based TITE-CRM, the most appealing feature of TITE-keyboard is that its dose transition rule can be tabulated before the trial begins; see Table \ref{table1} as the decision table with the target DLT rate $\phi=0.3$. To conduct the trial, there is no need for real-time model fitting, investigators only need to count the number of patients with DLTs (i.e., $\tilde{y}_j$), the number of patients with data pending (i.e., $\tilde{c}_j=\sum_{i=1}^{n_j}(1-\delta_i)$), the ``effective" number of patients without DLT (i.e., $\tilde{m}_j$),  and then use the decision table to determine the dose assignment for the next new cohort. Another feature of the TITE-keyboard design is that its decision table does not depend on the weighting scheme or the length of the DLT assessment window. Table \ref{table1} applies no matter which of the aforementioned three weighting schemes is used and no matter  the length of the assessment window. This is because the likelihood (\ref{approxlik}) only depends on $\tilde{m}_j$, which is the ``effective'' number of patients without DLT. Moreover, when all the pending DLT data become available, i.e., $\tilde{D}_j={D}_j$, the TITE-keyboard design reduces to the standard keyboard design in a seamless way.

For patient safety, we require that dose escalation is not allowed until at least 2 patients have completed the DLT assessment at the current dose level. In addition, we impose an overdose control/stopping rule: at any time during the trial, if any dose $j^\prime$ satisfies $\Pr(p_{j^\prime}>\phi\mid n_{j^\prime},\tilde{y}_{j^\prime})>\eta$, then that dose and any higher doses are regarded as overly toxic and should be eliminated from the trial, and the   dose is de-escalated to level $j^\prime-1$ for the next patients, where $\eta$ is the prespecified elimination cutoff, say $\eta=0.95$. If the lowest dose level is eliminated, the trial should be early terminated. Table \ref{table1} also reflects such safety and overdose control rules.

When dealing with  fast accrual or late-onset toxicities, the top concern is patient safety as the patients with pending outcome data who have not experienced DLT at the interim decision time may yet experience DLT late in the follow-up period. Any reasonable design that handles late-onset toxicity should take that fact into account in its decision making, which can be described by the monotonicity property.  Let $D^s_j$ denote the ``cross-sectional" interim data obtained by setting $ x_i = \tilde{x}_i $, i.e., treating the patients' temporary DLT outcomes at the interim time as their final DLT outcomes at the end of the assessment window. In other words, 
$D^s_j=(n_j,\tilde{y}_j)$. Let $a(D_j)=-1, 0$ and 1  respectively denote the decisions of dose de-escalation, retaining the current dose and dose escalation based on the data $D_j$.  

\noindent{\bf Definition} ({\it Monotonicity}) $\,\,$ A dose-finding design is monotonic if $a(D^o_j)\le a(D^s_j)$ for $j=1,\ldots,J$. 

Monotonicity  indicates that the decision of dose transition based on the observed data $D^o_j$ should be less aggressive than that based on $D^s_j$. This is a property that any reasonable design should obey to reflect that patients who have not experienced DLT by the interim decision time may yet experience  DLT late in the follow-up period. As shown in the Supplementary Material, the TITE-keyboard design has this property.

\begin{thm}\label{inconsistent}
The TITE-keyboard design is monotonic.
\end{thm}

Another finite-sample design property of practical importance is coherence.   \cite{cheung2005} originally defined coherence as a design property by which dose escalation (or de-escalation) is prohibited when the most recently treated patient experiences (or does not experience) toxicity. \cite{liu2015} extended that concept and defined two different types of coherence: short-memory coherence and long-memory coherence. They referred to the coherence proposed by \cite{cheung2005} as short-memory coherence because it concerns the observation from only the most recently treated patient, ignoring the observations from the patients who were previously treated. Long-memory coherence is defined as a design property by which dose escalation (or de-escalation) is prohibited when the observed toxicity rate among all accumulative patients treated at the current dose is larger (or smaller) than the target toxicity rate. From a practical viewpoint, long-memory coherence is more relevant because when clinicians determine whether a dose assignment is practically plausible, they almost always base their decision on the toxicity data that have accumulated from all patients, rather than only the single patient most recently treated at that dose. In practice, patients in phase I trials are very heterogeneous, therefore, the toxicity outcome from a single patient can be spurious. For example, suppose the target DLT rate $\phi=0.3$ and at the current dose, the most recently treated patient experienced DLT but none of the 7 patients previously treated at the same dose had DLT. As the overall observed DLT rate at the current dose is 1/8, escalating the dose should not be regarded as an inappropriate action, although it violates short-memory coherence.  It can be shown that the keyboard design is long-memory coherent. The proof of Theorem \ref{coherence} is given in the Supplementary Material.

\begin{thm}\label{coherence}
The TITE-keyboard design is long-memory coherent in the sense that the probability of escalating (or de-escalating) the dose if  the estimated toxicity rate $\tilde{p}_{j}=\tilde{y}_{j}/\tilde{n}_j$ at the current dose is greater (or less) than the target toxicity rate, that is, $\Pr(\mbox{Escalation}\mid\tilde{p}_{j}>\phi,D_{j}^o)=0$, and
$\Pr(\mbox{De-escalation}\mid\tilde{p}_{j}<\phi,D_{j}^o)=0$.
\end{thm}

In addition to monotonicity and coherence,  the TITE-keyboard design also possesses a desirable large-sample convergence property. The proof is provided in the Supplementary Material. 
\begin{thm}\label{convergence}
Given a finite assessment window, as the number of treated patients goes to infinity, the dose assignment of the TITE-keyboard design converges almost surely to the dose level $j$ with $p_j\in (\phi-\delta_1,\phi+\delta_2)$. 
\end{thm}

\subsection{Extension to mTPI and BOIN designs}
The proposed methodology can be directly applied to mTPI, another model-assisted design. The mTPI design specifies three dosing intervals: the proper dosing interval  ${\cal I}_0 = (\phi-\delta_1, \phi+\delta_2)$, underdosing interval ${\cal I}_1 = (0, \phi-\delta_1)$ and overdosing interval ${\cal I}_{-1} = (\phi+\delta_2, 1)$.
The dose escalation/de-escalation decision is based on the UPM of the three intervals:
\begin{equation}
{\rm UPM}_k(D_j) = \frac{\Pr(p_j \in {\cal I}_k|D_j)}{ \mathrm{length \,\, of\,\, }{\cal I}_k}= \frac{\int_{\mathcal{I}_{k}} {f}(p_{j}\mid D_{j}){\rm d}p_j}{\int_{\mathcal{I}_{k}}1{\rm d}p_{j}}, \qquad k=-1, 0, 1. \label{upm}
\end{equation}
The mTPI design assumes  the same beta-binomial model (\ref{betabinom}) and calculates the UPM$_k$'s based on the posterior of $p_j$  given by (\ref{bbpost}). The dose-assignment decision of mTPI is then $a(D_{j})=\underset{k=\{-1,0,1\}}{\arg\max}{\rm UMP}_{k}(D_{j})$; that is, making the decision that corresponds to the largest UPM. 

To apply the proposed methodology, the only change needed is to replace the complete data $D_j$, potentially unobserved due to late-onset toxicity or fast accrual,  by the ``effective" data $\tilde{D}_j$ in (\ref{upm}) when calculating the UPMs. Once the UPMs are determined, the decision rules remain the same. We refer to the resulting design as the TITE-mTPI. Table S1 in the Supplementary Material shows the dose escalation/de-escalation rule for the TITE-mTPI design.

Last, we briefly discuss how to apply our methodology to the BOIN design. The BOIN design has a simpler and more transparent decision rule than the mTPI and keyboard designs. It does not require calculating the posterior distribution of $p_j$ and enumerating all possible outcomes for $(n_j, y_j)$. BOIN makes the decision of dose escalation/de-escalation by comparing the observed DLT rate $\hat{p}_j=y_j/n_j$ with a pair of dose escalation and de-escalation boundaries $(\lambda_e, \lambda_d)$ that are optimized to minimize the probability of incorrect dose-assignment decisions, assuming $y_j \sim {\rm Binom}(n_j, p_j)$. If $\hat{p}_j\le\lambda_e$, escalate the dose; if $\hat{p}_j \ge \lambda_d$, de-escalate the dose; otherwise, i.e., $\lambda_e<\hat{p}_j <\lambda_d$, stay at the same dose. \cite{liu2015}  provided a closed-form formula for $(\lambda_e, \lambda_d)$, and showed that the dose escalation/de-escalation rule of the BOIN is equivalent to using the frequentist likelihood ratio test statistic, or Bayesian factor,  to guide the dose transition. That is, BOIN has both frequentist and Bayesian interpretations.

 In the presence of pending DLT data due to fast accrual or late-onset toxicity, BOIN cannot be used because $\hat{p}_j$ is not available as $y_j$ is unknown. To address this issue, utilizing our approximation with the ``effective" binomial data $(\tilde{n}_j, \tilde{y}_j)$, we can simply use $\tilde{p}_j = \tilde{y}_j/\tilde{n}_j$ to replace $\hat{p}_j$ as the maximum likelihood estimate of $p_j$,  and then make the decision of dose escalation/de-escalation in the same way as in the original BOIN. Due to space limitation, the operating characteristics and theoretical properties of this approach are  not studied here.

\subsection{Software}
To facilitate the use of the TITE-keyboard and TITE-mTPI designs, we have developed graphical user interface-based software that allows users to generate the dose-assignment decision table, conduct simulations, obtain the operating characteristics of the design, and generate a trial design template for protocol preparation. The software will be freely available at \url{http://www.trialdesign.org}.

\section{Trial illustration}
\label{Sec:Trial}
We apply the proposed TITE-keyboard design to the melanoma trial described in the Introduction. The target toxicity probability $\phi=0.3$, assessment window $\tau=3$ months, and patients are enrolled at the rate of 2 patients per month. The maximum sample size is 21 patients. Patients are treated in cohorts of 3, starting from the lowest dose level.

\begin{figure}
\begin{center}
\includegraphics[width=4.5in,height=3in]{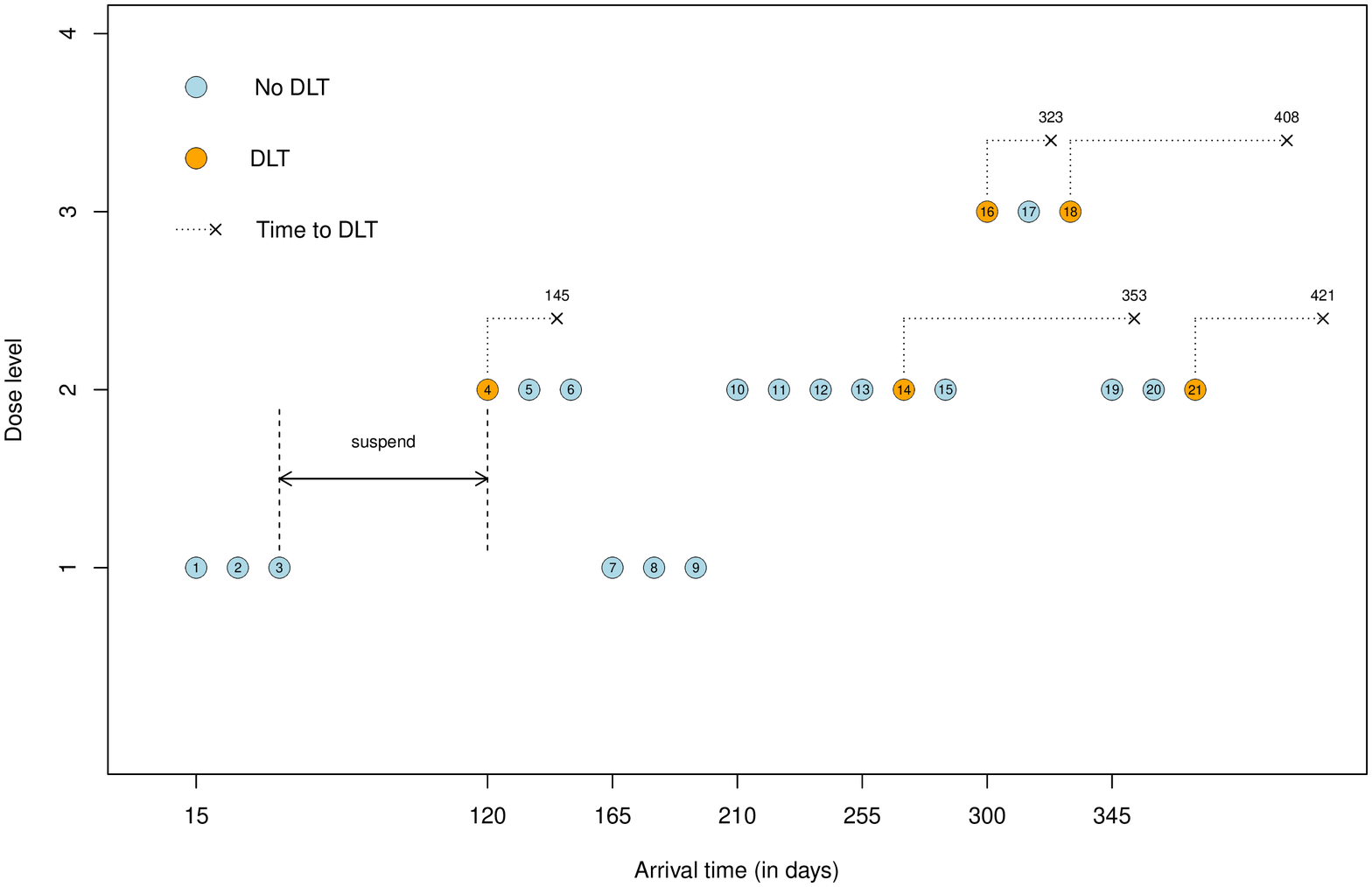}
\end{center}
\caption{Hypothetical phase I clinical trial using the TITE-Keyboard design. The target toxicity rate is 0.3, and the toxicity assessment window is 3 months. Patients are treated in cohort sizes of 3, and the accrual rate is one patient every 15 days. 
The number above the ``x'' indicates the time when DLT occurs.}
\label{trialex}
\end{figure}

Given the target toxicity rate of 0.3, 
Table \ref{table1} shows the TITE-keyboard decision rule with a cohort size of 3. Figure \ref{trialex} displays the whole trial procedure using the TITE-keyboard design and the uniform scheme $w_i^{\rm u}$. Patients in the first cohort were treated at the lowest dose level. Since no DLT was observed and the TITE-keyboard design requires at least two finished patients before dose escalation, the trial was suspended until day 120 when the first two patients had completed the assessment. Following the TITE-keyboard decision rule, the second cohort was treated at dose level 2. On the arrival of the third cohort (day 165), one DLT had occurred on day 145 among the second cohort,  and the follow-up proportions for patients 5 and 6 were 1/3 and  1/6, respectively, leading to $\tilde{m}_j=0.5$. 
According to Table \ref{table1}, dose de-escalation was needed. 
Had we treated the two pending outcomes as non-DLTs,  the keyboard design based on $(n_j,y_j)=(3,1)$ would have recommended retaining the current dose for the next cohort of patients. However, since the two patients with pending outcome data had been followed for only a short period,  there was greater uncertainty regarding the toxicity probabilities of dose level 2 and it was preferable to be conservative. The TITE-keyboard automatically took into account such uncertainty and de-escalated the dose to the first level for treating the third cohort. When patient 10 arrived, no DLT was observed among the 6 patients at that dose level, thus the forth cohort received dose level 2. On day 255, 6 patients had been treated at level 2, with 3 finished patients. The observed data at level 2 were $(n_j,y_j,c_j,\tilde{m}_j)=(6,1,3,3)$, which were less than the dose-escalation boundary. As a result, the fifth cohort was still treated at dose level 2. On day 300, the observed data at that level were updated to be $(n_j,y_j,c_j,\tilde{m}_j)=(9,1,5,5.5)$, and the dose for patients 15 to 18 was escalated to level 3. As one DLT was observed at level 3 before the arrival of patient 19, patients in the last cohort were assigned to dose level 2. Based on TITE-keyboard, the total trial duration was 14 months. In contrast, the trial would have run about 31.5 months if we had applied standard adaptive designs that require full DLT assessment before enrolling each new cohort. 
At the end of the trial, a total of 12 patients had been treated at dose level 2, and 3 DLTs were observed. The estimated toxicity rate at dose level 2 was 0.25, thus it was selected as the MTD. 

To assess the accuracy of the approximated likelihood  function (\ref{approxlik}), we provide the decisions by the TITE-keyboard design respectively with  the true likelihood (\ref{likelihood})
and the approximated likelihood (\ref{approxlik}) in Table S2 in the Supplementary Material. As expected, the posterior model probabilities calculated based on the true and approximated likelihoods are almost identical, thus, the decisions made based on the approximated likelihood are consistent with those based on the true likelihood.

\section{Numerical studies}
\label{Sec:Simu}
\subsection{Fixed scenarios}
\label{fixsim}
We perform simulation studies to examine the operating characteristics of the proposed TITE-keyboard and TITE-mTPI designs. We consider six dose levels, with the target toxicity probability $\phi=0.30$. The toxicity assessment window is $\tau=3$ months.
A maximum of 36 patients will be recruited in cohorts of 3, with the accrual rate of 2 patients per month. Six toxicity scenarios with different locations of the MTD and various shapes of dose--toxicity curves are considered; see Table \ref{fixres}. The time-to-toxicity outcomes are simulated from Weibull distributions by controlling that 50\% of the toxicity events  occur in the latter half of the assessment window $(\tau/2,\tau)$ \citep{liu2013}. Under each scenario, we compare the proposed TITE-keyboard and TITE-mTPI designs with the R6 design, TITE-CRM, and the standard CRM based on 10,000 simulated trials. The uniform weighting scheme  is adopted for the TITE-CRM, TITE-keyboard and TITE-mTPI designs.
For TITE-CRM and CRM, the power model is utilized, where the skeleton is chosen based on  the method of  \cite{lee2009}, with the initial MTD guess being dose level 3 and the half-width of the indifference interval being 0.06. The (standard) CRM cannot directly handle late-onset toxicity. To implement the CRM, we suspend accrual after treating each cohort of patients until all pending data are fully observed before enrolling the next cohort of patients.  The CRM serves as the benchmark to assess the performance of the other designs.  For fair comparison, in the TITE-keyboard, TITE-mTPI and TITE-CRM designs, we apply the same safety rule that dose escalation is not allowed until two patients at the current dose level have finished the DLT assessment. When the R6 design  stops the trial early (e.g., when 2 DLTs are observed among 2 patients) before the exhaustion of 36 patients, the remaining patients are treated at the selected ``MTD'' as the cohort expansion, such that the sample sizes of the five designs are matched.

Table \ref{fixres} summarizes the simulation results, including the dose selection percentage; the  percentage  of patients treated at each dose; the average trial duration; the early stopping percentage; the risk of poor allocation, defined as the percentage of
simulated trials allocating fewer than 6 patients to the MTD;  and the risk of overdosing, defined as the percentage of
simulated trials that treat more than half of the patients at doses above the MTD. 
In terms of MTD selection and patient allocation, the TITE-CRM, TITE-mTPI and TITE-keyboard designs yield performances generally comparable to those of the CRM, while the R6 design performs the worst as it inherits the drawbacks of the 3+3 design. Compared to the CRM, 
the TITE-CRM, TITE-mTPI and TITE-keyboard designs dramatically shorten the trial duration by approximately 22 months. In scenarios 1 and 2, TITE-CRM requires a slightly shorter trial duration than TITE-keyboard because it is more likely to stop the trial early. 
R6 requires repeated suspension of accrual until  6 patients have been treated at the current dose level. As a result, it generally has a longer trial duration than the other three time-to-event methods.  

When dealing with late-onset toxicities, it is not desirable to be overly conservative and retain low doses too long because that allocates too many patients to subtherapeutic doses. This early-settlement problem is quantified by the risk of poor allocation. Among the considered designs, the TITE-keyboard and TITE-CRM have the lowest risk of poor allocation, indicating that these two designs can quickly recover from the false settlement and allocate patients to the appropriate doses. TITE-mTPI is more likely to become stuck at incorrect doses than TITE-keyboard and TITE-CRM, with a higher risk of poor allocation.  The R6 design performs the worst since it is  excessively conservative and would not escalate the dose until 6 patients have been treated at the current dose level. 
When handling late-onset toxicities, it is also not desirable to be overly aggressive and treat a large percentage of patients at overly toxic doses. We measure this design behavior using the risk of overdosing.  According to Table \ref{fixres}, the TITE-keyboard design is safer than the TITE-CRM and TITE-mTPI designs, demonstrating the lowest risk of overdosing patients among the three designs.

To investigate the accuracy of our proposed approximation procedure, we implement the TITE-keyboard$^{\rm t}$ design that is based on the true likelihood function (\ref{likelihood}), where the superscript  ``${\rm t}$'' represents the true likelihood.  
As shown in Table \ref{fixres}, the operating characteristics of the TITE-keyboard design are essentially identical to those of the 
TITE-keyboard$^{\rm t}$ design. To gain more insight, we record the percentage of discrepancy in dose assignment between the two designs across 10,000 simulated trials, which is merely about  1.3\% on average for the considered scenarios. These results verify the high accuracy of the approximated likelihood function.

\subsection{Sensitivity analysis on different calculation schemes of $w_i$}
\begin{figure}
\begin{center}
\includegraphics[width=4.5in,height=3.5in]{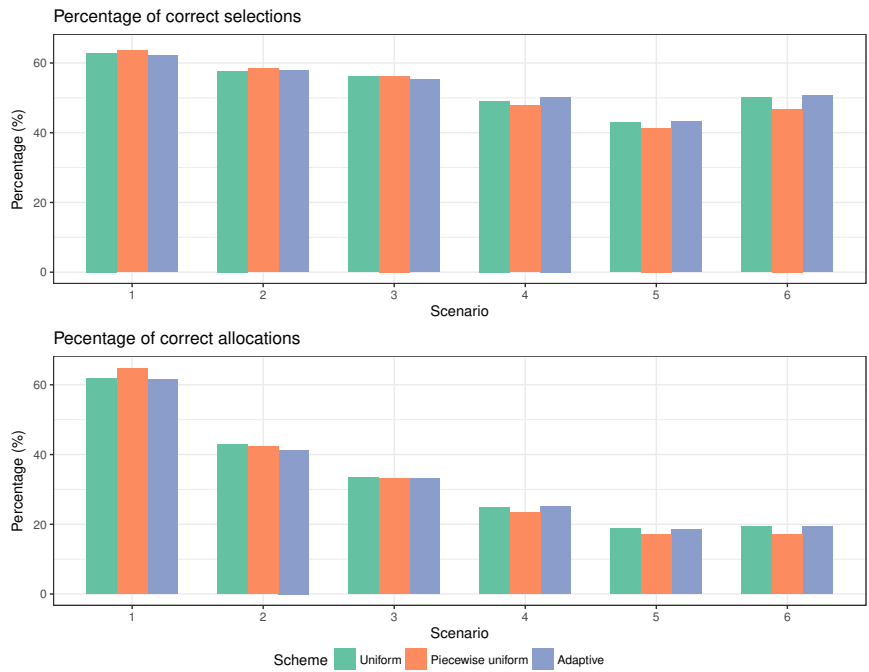}
\end{center}
\caption{Sensitivity analysis of different  schemes for the weight $w_i=\Pr(t_{i}<u_{i}\mid x_{i}=1)$ based on the six fixed scenarios in Table \ref{fixres}. Three different schemes are considered. Uniform: $t_i\mid x_i=1\sim {\rm Unif}(0,\tau)$; Piecewise uniform: $t_i\mid x_i=1\sim \upsilon_1{\rm Unif}(0,\tau/3)+\upsilon_2{\rm Unif}(\tau/3,2\tau/3)+\upsilon_3{\rm Unif}(2\tau/3,\tau)$; Adaptive: $w_i$ is adaptively estimated based on the observed data and the prior $t_i\mid x_i=1\sim \tau\times{\rm Beta}(\lambda,\gamma)$, $\lambda,\gamma \sim {\rm Gamma}(0.5,0.5)$.}
\label{sens}
\end{figure}

To assess the robustness of TITE-keyboard to different weighting schemes for $w_i$, we compare the performance of TITE-keyboard based on the uniform scheme with those based on the piecewise uniform and adaptive schemes. 
For the piecewise  uniform scheme, we choose $(\upsilon_1,\upsilon_2,\upsilon_3)=(1/6,2/6,3/6)$ so that the prior information indicates that half of the toxicities may occur in the final part of the assessment window. For the adaptive scheme, we take independent ${\rm Gamma}(0.5,0.5)$ prior distributions for $\lambda$ and $\gamma$. We respectively refer to the TITE-keyboard designs based on these three calculation schemes as TITE-keyboard$^{\rm u}$, TITE-keyboard$^{\rm p}$, and TITE-keyboard$^{\rm a}$, and 
compare their percentages of correct selections and percentages of correct allocations in 
 Figure \ref{sens}.  
The performance of TITE-keyboard$^{\rm a}$ is almost the same as that of TITE-keyboard$^{\rm u}$. There are some very slight differences between the operating characteristics of TITE-keyboard$^{\rm p}$ and TITE-keyboard$^{\rm u}$. When more late-onset prior information (i.e., the piecewise uniform scheme) is incorporated, the TITE-keyboard$^{\rm p}$ design tends to be more conservative compared to that using the uniform scheme: if the MTD lies in the lower dose region (scenarios 1 and 2), the performance of TITE-keyboard$^{\rm p}$ is better than that of TITE-keyboard$^{\rm u}$; and vice versa,  TITE-keyboard$^{\rm u}$ outperforms TITE-keyboard$^{\rm p}$ in scenarios 5 and 6, where high dose levels are the MTD. Nonetheless, the differences between the operating characteristics of TITE-keyboard$^{\rm p}$ and TITE-keyboard$^{\rm u}$ are very minor across the six considered scenarios, showing that the TITE-keyboard design is not sensitive to the weighting schemes of $w_i$.

\subsection{Random scenarios}

To ensure that the simulation results represent the general performance of the designs, we conducted a large-scale simulation study
to compare the proposed TITE-keyboard, TITE-mTPI designs with R6 and TITE-CRM based on randomly generated dose--toxicity scenarios. 
We considered 12 configurations that cover various target toxicity rates, numbers of cohorts, cohort sizes, distributions of time-to-toxicity outcomes, percentages of toxicity occurring in the latter half of the assessment window $(\tau/2,\tau)$, and accrual rates. Details of the configurations are provided in Table S3 in the Supplementary Material. Under each configuration, we randomly generated 50,000 random dose--toxicity scenarios based on the procedure described in \cite{zhou2018a}. 
The setup of the comparative designs is the same as that described in Section \ref{fixsim}.

\begin{figure}
\begin{center}
\includegraphics[width=5in,height=7in]{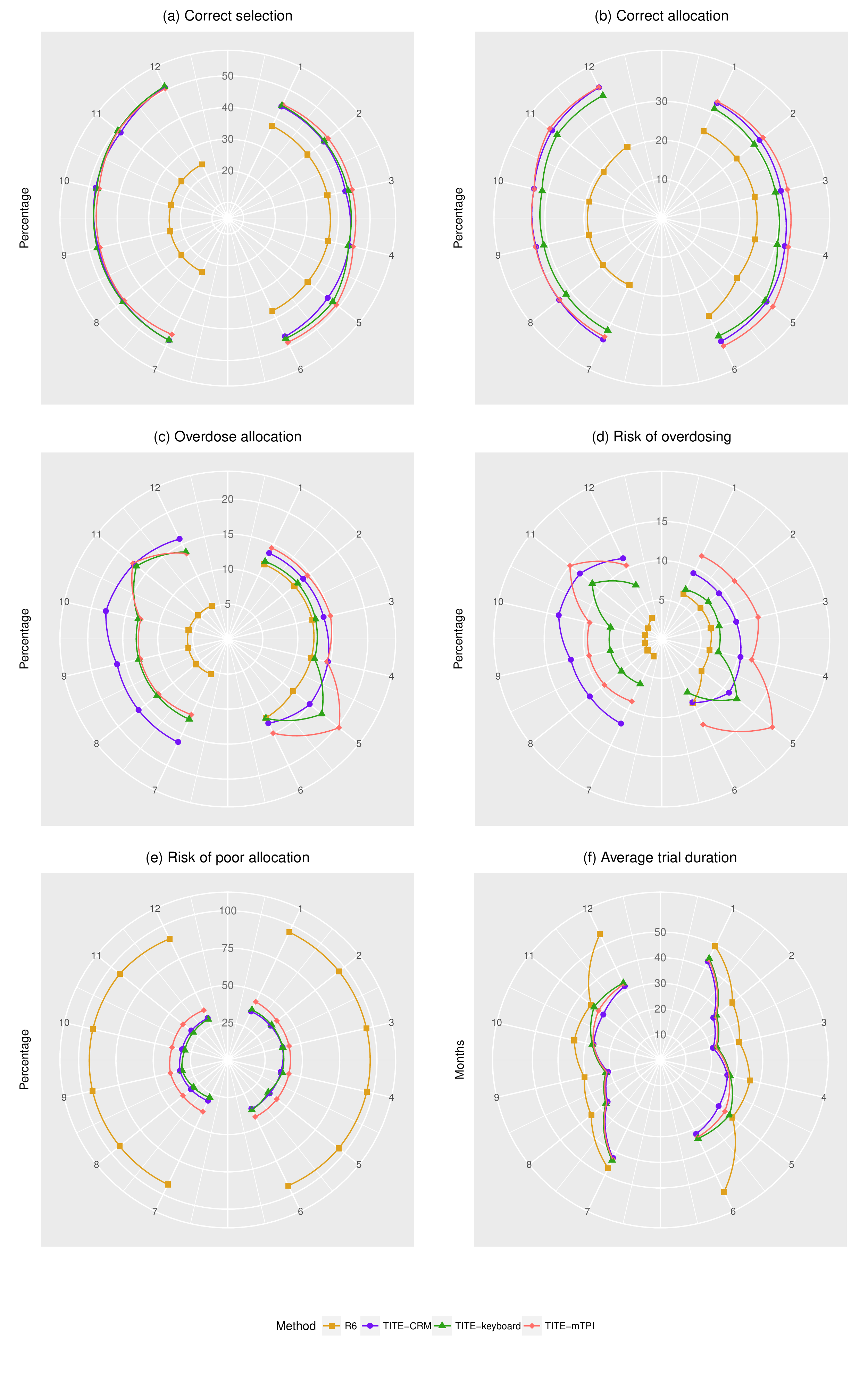}
\end{center}
\caption{Simulation results based on 50000 randomly generated scenarios. The target toxicity probability in scenarios 1--6 is 0.2, while that in scenarios 7--12 is 0.3.}
\label{randres1}
\end{figure}

Figure \ref{randres1}   shows the simulation results, which  are generally consistent with those obtained in the fixed scenarios. TITE-keyboard, TITE-mTPI and TITE-CRM have similar accuracy when identifying the MTD, and they uniformly outperform the R6 design.  TITE-keyboard is safer and easier to implement than TITE-CRM. Between the two model-assisted designs, TITE-keyboard has substantially lower risk of overdosing patients and having poor allocation than TITE-mTPI and thus is preferable for practical use.  By comparing the operating characteristics of TITE-keyboard across the 12 simulation configurations, we additionally found that the performance of TITE-keyboard  is robust to the  late-onset toxicity profile, including the ratio of the assessment window and the patient inter-arrival time as well as the distribution of time-to-toxicity outcomes. 

\section{Concluding remarks}
\label{Sec:Con}
We have proposed general methodology to allow model-assisted trial designs to handle  late-onset toxicity and fast accrual. We have formulated a new likelihood-based approach to account for pending DLT data and have derived a novel approximation of the observed likelihood that enables all existing model-assisted designs to accommodate pending data in a seamless way without destroying their simplicity.  In particular, we have proposed the TITE-keyboard design, which has been demonstrated to possess desirable finite-sample and large-sample properties: monotonicity, coherence and consistency. The TITE-keyboard design is simple and easy to implement, yet has superior performance that is comparable to that of the more complicated TITE-CRM design. 

Conducting clinical trials is a complicated, multidisciplinary effort, involving clinicians, a medical support team and statisticians. When determining a trial design to use in practice, transparency and simplicity are viewed as important as statistical properties. The proposed designs allow users to tabulate the dose escalation and de-escalation rules before the trial begins. To conduct the trial, there is no complicated model fitting and calculation, the investigators only need to look up the decision table to make the decision of dose escalation and de-escalation, which greatly simplifies the implementation of the design. Importantly, such simplicity does not sacrifice performance. Thus, the proposed TITE-keyboard design has great potential for being adopted by practitioners to accelerate drug development. 

To be conservative, we adopt an ad hoc rule to prohibit dose escalation and suspend patient accrual if fewer than two patients at the current dose level have been completely followed. As another approach, we can only allow for dose escalation if $\Pr(p_j<\phi \mid n_j,\tilde{y}_j)$ is greater than some prespecified threshold. However, due to the small sample size and the nature of the binomial likelihood, these two rules tend to produce similar operating characteristics.

\section{Supplementary Materials}
This supplementary material consists of the proofs of Theorems 1--4 and other technical details given in the paper.

\bibliographystyle{plainnat}

\begin{thebibliography}{99}

\bibitem[Babb et al.(1998)]{babb1998}
{Babb, J., Rogatko, A., and Zacks, S.} (1998).
Cancer phase I clinical trials: efficient dose escalation with overdose control.
\emph{Statistics in Medicine} {\bf 17}, 1103--1120.


\bibitem[Barlow and Brunk(1972)]{barlow1972}
Barlow, R. E. and Brunk, H. D. (1972). The isotonic regression problem and its dual. {\it Journal of the American Statistical Association} {\bf 67}, 140--147.

%\bibitem[Bekele et al.(2008)]{bekele2008}
%{Bekele, B. N., Ji, Y., Shen, Y., and Thall, P. F.} (2008). Monitoring late-onset toxicities in phase I trials using predicted risks. {\it Biostatistics} {\bf 9}, 442--457.
%
 
%\bibitem[Clertant and  O'Quigley(2017)]{clertant2017}
%Clertant, M. and O'Quigley, J. (2017). Semiparametric dose finding methods. {\it Journal of the Royal Statistical Society: Series B (Statistical Methodology)} {\bf 79}, 1487--1508.

\bibitem[Cheung(2005)]{cheung2005}
{Cheung, Y. K.} (2005). Coherence principles in dose-finding studies.
{\it Biometrika} {\bf 92}, 863--873.

\bibitem[Cheung and Chappell(2000)]{cheung2000}
{Cheung, Y. K. and Chappell, R.} (2000).
Sequential designs for phase I clinical trials with late-onset toxicities.
{\it Biometrics} {\bf 56}, 1177--1182.
 
 \bibitem[Dilts et al.(2008)]{Dilts08}
Dilts, D. M., Sandler, A. B. , Cheng, S., Crites, J., Ferranti, L., Wu, A., Bookman, M. A., Thomas, J. P.  and Ostroff J. (2008). Accrual to clinical trials at selected comprehensive cancer centers. {\it Journal of Clinical Oncology}   {\bf 26}, 6543--6543 

 \bibitem[Guo et al.(2017)]{Guoji17}
Guo, W., Wang, S. J., Yang, S., Lynn, H., Ji, Y. (2017). A Bayesian interval dose-finding design addressing Ockham?s razor: MTPI-2. {\it Contemporary Clinical Trials} {\bf 58}, 23--33.


\bibitem[Ivanova et al.(2007)%
Ivanova, Flournoy and Chung]{ivanova2007}
{Ivanova, A., Flournoy, N., and Chung, Y.} (2007).
Cumulative cohort design for dose finding.
\emph{ Journal of Statistical Planning and Inference}
 {\bf 137},
 2316--2327.
 
\bibitem[Ji et al.(2010)]{ji2010} 
Ji, Y., Liu, P., Li, Y., and Nebiyou Bekele, B. (2010). A modified toxicity probability interval method for dose-finding trials. {\it Clinical Trials} {\bf 7}, 653--663.
 
\bibitem[Jin et al.(2014)]{jin2014}
Jin, I. H., Liu, S., Thall, P. F., and Yuan, Y. (2014). Using data augmentation to facilitate conduct of phase I--II clinical trials with delayed outcomes. {\it Journal of the American Statistical Association} {\bf 109}, 525--536.

\bibitem[June et al.(2017)]{june2017}
June CH, Warshauer JT, Bluestone JA. (2017) Is autoimmunity the Achilles' heel of cancer immunotherapy? {\it Nature Medicine} {\bf 23},  540-547.


\bibitem[Kairalla et al.(2012)]{kairalla2012}
Kairalla, J. A., Coffey, C. S., Thomann, M. A., and Muller, K. E. (2012). Adaptive trial designs: a review of barriers and opportunities. {\it Trials} {\bf 13}, 145.


\bibitem[Lee and Cheung(2009)]{lee2009}
Lee, S. M. and Cheung, Y. K. (2009). Model calibration in the continual reassessment method. {\it Clinical Trials} {\bf 6}, 227--238.

\bibitem[Lin and Yin(2017a)]{lin2017a}
Lin, R. and Yin, G. (2017a). Nonparametric overdose control with late-onset toxicity in phase I clinical trials. {\it Biostatistics} {\bf 18}, 180--194.


\bibitem[Lin and Yin(2017b)]{lin2016}
Lin, R. and Yin, G. (2017b). Bayesian optimal interval design for dose finding in drug-combination trials.
{\it Statistical Methods in Medical Research} {\bf 26}, 2155--216.

\bibitem[Lin and Yin(2017c)]{lin2017}
Lin, R. and Yin, G. (2017c). STEIN: A simple toxicity and efficacy interval design for seamless phase I/II clinical trials. {\it Statistics in Medicine} {\bf 36}, 4106--4120.




\bibitem[Liu et al.(2013)]{liu2013}
{Liu, S., Yin, G., and Yuan, Y.} (2013). Bayesian data augmentation dose finding with continual reassessment method and delayed toxicity. {\it The Annals of Applied Atatistics} {\bf 7}, 2138--2156.


\bibitem[Liu and Yuan(2015)%
Liu and Yuan]{liu2015}
{Liu, S. and Yuan, Y.} (2015).
Bayesian optimal interval designs for phase I clinical trials.
\emph{Journal of Royal Statistical Society: Series C (Applied Statistics)}
{\bf 64}, 507--523.

 \bibitem[Mauguen et al.(2011)]{mauguen2011}
{Mauguen, A., Le Deley, M. C., and Zohar, S.} (2011). Dose-finding approach for dose escalation with overdose control considering incomplete observations. {\it Statistics in Medicine} {\bf 30}, 1584--1594.

 \bibitem[Mu et al.(2018)]{mu2018}
Mu, R., Yuan, Y., Xu, J., Mandrekar, S. J., and Yin, J. Y. (2018). gBOIN: A unified model-assisted phase I trial design accounting for toxicity grades, binary or continuous
endpoint. {\it Journal of the Royal Statistical Society: Series C (Applied Statistics)}; in press.

\bibitem[Muler et al.(2004)]{muler2004}
Muler, J. H., McGinn, C. J., Normolle, D., Lawrence, T., Brown, D., Hejna, G., and Zalupski, M. M. (2004). Phase I trial using a time-to-event continual reassessment strategy for dose escalation of cisplatin combined with gemcitabine and radiation therapy in pancreatic cancer. {\it Journal of Clinical Oncology} {\bf 22}, 238--243.

\bibitem[Normolle and Lawrence(2006)]{normolle2006}
Normolle, D. and Lawrence, T. (2006). Designing dose-escalation trials with late-onset toxicities using the time-to-event continual reassessment method. {\it Journal of Clinical Oncology} {\bf 24}, 4426--4433.


\bibitem[O'Quigley et al.(1990)%
O'Quigley, Pepe  and Fisher]{quigley1990}
{O'Quigley, J., Pepe, M., and Fisher, L.} (1990).
Continual reassessment method: a practical design for phase 1 clinical trials in cancer.
\emph{Biometrics}
 {\bf 46},
 33--48.
 
 \bibitem[Pan et al.(2018)]{pan2018}
Pan, H., Lin, R., and Yuan, Y. (2018). Statistical properties of the keyboard design with extension to drug-combination trials.   Manuscript.


 \bibitem[Postel-Vinay(2011)]{postel2011}
 {Postel-Vinay, S., Gomez-Roca, C., Molife, L. R., Anghan, B., Levy, A., Judson, I.,
 Bono, J. D.,  Soria, JC., Kaye, S.,
 and Paoletti, X.} (2011). Phase I trials of molecularly targeted agents: should we pay more attention to late toxicities?
 {\it Journal of Clinical Oncology} {\bf 29}, 1728--1735.
 
 
\bibitem[Skolnik et al.(2008)]{skolnik2008} 
Skolnik, J. M., Barrett, J. S., Jayaraman, B., Patel, D., and Adamson, P. C. (2008). Shortening the timeline of pediatric phase I trials: the rolling six design. {\it Journal of Clinical Oncology} {\bf 26}, 190--195.

\bibitem[Storer(1989)%
Storer]{storer1989}
{Storer, B. E.} (1989).
Design and analysis of phase I clinical trials.
\emph{ Biometrics}
 {\bf 45},
 925--937.
 
 \bibitem[Weber et al.(2015)]{weber2015}
 Weber, J. S., Yang, J. C., Atkins, M. B., et al. (2015). Toxicities of immunotherapy for the practitioner. {\it Journal of Clinical Oncology} {\bf 33},  2092--2099.

\bibitem[Yan et al.(2017)]{yan2017}
Yan, F., Mandrekar, S. J., and Yuan, Y. (2017). Keyboard: a novel Bayesian toxicity probability interval design for phase I clinical trials. {\it Clinical Cancer Research} {\bf 23}, 3994--4003.

\bibitem[Yin(2012)]{yinbook2012}
{Yin, G.} (2012). Clinical Trial Design: Bayesian and Frequentist Adaptive Methods. New
York: John Willey $\&$ Sons, Inc.

 

%\bibitem[Yin et al.(2013)]{yin2013}
%{Yin, G., Zheng, S., and Xu, J.} (2013). Fractional dose-finding methods with late-onset toxicity in phase I clinical trials. {\it Journal of Biopharmaceutical Statistics} {\bf 23}, 856--870.

\bibitem[Yuan and Yin(2009)]{yuan2009}
Yuan, Y. and Yin, G. (2009). Bayesian dose finding by jointly modelling toxicity and efficacy as time-to-event outcomes. {\it Journal of the Royal Statistical Society: Series C (Applied Statistics)} {\bf 58}, 719--736.

\bibitem[Yuan and Yin(2011)]{yuan2011}
{Yuan, Y. and Yin, G.} (2011). Robust EM continual reassessment method in oncology dose finding.
{\it Journal of the American Statistical Association} {\bf 106}, 818--831.

 
\bibitem[Yuan et al.(2016)]{yuan2016}
Yuan, Y., Hess, K. R., Hilsenbeck, S. G., and Gilbert, M. R. (2016). Bayesian optimal interval design: a simple and well-performing design for phase I oncology trials. {\it Clinical Cancer Research} {\bf 22}, 4291--4301.
 
\bibitem[Zhang and Yuan(2016)]{zhang2016} 
Zhang, L. and Yuan, Y. (2016). A practical Bayesian design to identify the maximum tolerated dose contour for drug combination trials. {\it Statistics in Medicine} {\bf 35}, 4924--4936.

\bibitem[Zhao et al.(2011)]{zhao2011}
Zhao, L., Lee, J., Mody, R., and Braun, T. M. (2011). The superiority of the time-to-event continual reassessment method to the rolling six design in pediatric oncology Phase I trials. {\it Clinical Trials} {\bf 8}, 361--369.

 \bibitem[Zhou et al.(2018a)]{zhou2018a}
Zhou, H., Murray, T., Pan, H., and Yuan, Y. (2018a) Comparative review of model-assisted designs for phase I clinical trials. {\it Statistics in  Medicine}; in press.

 \bibitem[Zhou et al.(2018b)]{zhou2018b}
Zhou, H., Yuan, Y., and Nie, L. (2018b) Accuracy, safety and reliability of novel Bayesian phase I trial designs. {\it Clinical  Cancer Research}; in press.

 
\end{thebibliography}

\newpage
\clearpage

\begin{table}
\begin{center}
\caption{Dose escalation and de-escalation boundaries for TITE-keyboard with a target DLT rate of 0.3 and cohort size of 3, up to 12 patients. \label{table1}}
 \tabcolsep 0.15in
\begin{tabular}{lllccc}
\hline 
$n_{j}$ & $\tilde{y}_{j}$ & $\tilde{c}_{j}$ & Escalation & Stay & De-escalation\tabularnewline
\hline 
3 & 0 & $\leq1$ & Y &  & \tabularnewline
3 & 0 & $\geq2$ & \multicolumn{3}{c}{Suspend accrual}\tabularnewline
3 & 1 & 0 &  & Y & \tabularnewline
3 & 1 & $1\leq \tilde{c}_{j}\leq2$ &  & $\tilde{m}_{j}>1.88$ & $\tilde{m}_{j}\leq1.88$\tabularnewline
3 & $2$ & $\leq1$ &  &  & Y\tabularnewline
3 & $ 3 $ & $0$ &  &  & Y\&Eliminate\tabularnewline
6 & 0 & $\leq6$ & Y &  & \tabularnewline
6 & 1 & $\leq1$ & Y &  & \tabularnewline
6 & 1 & $2\leq \tilde{c}_{j}\leq3$ & $\tilde{m}_{j}\geq3.07$ & $\tilde{m}_{j}<3.07$ & \tabularnewline
6 & 1 & $4\leq \tilde{c}_{j}\leq5$ & $\tilde{m}_{j}\geq3.07$ & $1.88<\tilde{m}_{j}>3.07$ & $\tilde{m}_{j}\leq1.88$\tabularnewline
6 & 2 & 0 &  & Y & \tabularnewline
6 & 2 & $1\leq \tilde{c}_{j}\leq4$ &  & $\tilde{m}_{j}>3.75$ & $\tilde{m}_{j}\leq3.75$\tabularnewline
6 & $3$ & $\leq3$ &  &  & Y\tabularnewline
6 & $4$ & $\leq2$ &  &  & Y\&Eliminate\tabularnewline
9 & 0 & $\leq9$ & Y &  & \tabularnewline
9 & 1 & $\leq4$ & Y &  & \tabularnewline
9 & 1 & $5\leq \tilde{c}_{j}\leq6$ & $\tilde{m}_{j}\geq3.07$ & $\tilde{m}_{j}<3.07$ & \tabularnewline
9 & 1 & $7\leq \tilde{c}_{j}\leq8$ & $\tilde{m}_{j}\leq3.07$ & $1.88<\tilde{m_{j}}<3.08$ & $\tilde{m}_{j}\leq1.88$\tabularnewline
9 & 2 & 0 &  & Y & \tabularnewline
9 & 2 & $1\leq \tilde{c}_{j}\leq3$ & $\tilde{m}_{j}\geq6.15$ & $\tilde{m}_{j}<6.15$ & \tabularnewline
9 & 2 & $4\leq \tilde{c}_{j}\leq7$ & $\tilde{m}_{j}\geq6.15$ & $3.75<\tilde{m}_{j}<6.15$ & $\tilde{m}_{j}\leq3.75$\tabularnewline
9 & 3 & 0 &  & Y & \tabularnewline
9 & 3 & $1\leq \tilde{c}_{j}\leq6$ &  & $\tilde{m}_{j}>5.63$ & $\tilde{m}_{j}\leq5.63$\tabularnewline
9 & $4$ & $\leq5$ &  &  & Y\tabularnewline
9 & $5$ & $\leq4$ &  &  & Y\&Eliminate \tabularnewline
12 & 0 & $\leq12$ & Y &  & \tabularnewline
12 & 1 & $\leq7$ & Y &  & \tabularnewline
12 & 1 & $8\leq \tilde{c}_{j}\leq9$ & $\tilde{m}_{j}\geq3.07$ & $\tilde{m}_{j}<3.07$ & \tabularnewline
12 & 1 & $10\leq \tilde{c}_{j}\leq11$ & $\tilde{m}_{j}\geq3.07$ & $1.88<\tilde{m_{j}}<3.08$ & $\tilde{m}_{j}\leq1.88$\tabularnewline
12 & 2 & $\leq3$ & Y &  & \tabularnewline
12 & 2 & $4\leq \tilde{c}_{j}\leq6$ & $\tilde{m}_{j}\geq6.15$ & $\tilde{m}_{j}<6.15$ & \tabularnewline
12 & 2 & $7\leq \tilde{c}_{j}\leq10$ & $\tilde{m}_{j}\geq6.15$ & $3.75<\tilde{m}_{j}<6.15$ & $\tilde{m}_{j}\leq3.75$\tabularnewline
12 & 3 & $\leq3$ &  & Y & \tabularnewline
12 & 3 & $4\leq \tilde{c}_{j}\leq9$ &  & $\tilde{m}_{j}>5.63$ & $\tilde{m}_{j}\leq5.63$\tabularnewline
12 & 4 & 0 &  & Y & \tabularnewline
12 & 4 & $1\leq \tilde{c}_{j}\leq8$ &  & $\tilde{m}_{j}>7.50$ & $\tilde{m}_{j}\leq7.50$\tabularnewline
12 & 5,6 & $\leq7$ &  &  & Y\tabularnewline
12 & 7 & $\leq5$ &  &  & Y\&Eliminate\tabularnewline
\hline 
\end{tabular}
\vspace{0.5in}
\end{center}
{\footnotesize Note: $n_j$ is the number of patients at dose level $j$, $\tilde{y}_j$ is the number of DLTs observed  by the decision time, $\tilde{c}_j=\sum_{i=1}^{n_j}(1-\delta_i)$ is the number of patients who have data pending, and $\tilde{m}_j$ is the effective number of patients without any DLT. Dose escalation is not allowed if fewer than 2 patients at dose level $j$ have  finished the assessment. ``Y" means Yes.} 
\end{table}

\begin{table}
\begin{center}
\caption{Simulation results with sample size of 36 and cohort size of 3. The assessment window is 3 months and the accrual rate is 2 patients per month. The MTD is in boldface. \label{fixres}}
 \tabcolsep 0.05in
 \renewcommand\arraystretch{1}
 {\footnotesize
\begin{tabular}{llcccccccccc}
\hline 
\multirow{2}{*}{Methods} &  & \multicolumn{6}{c}{Dose level} & Duration & \multirow{2}{*}{Stop\%} & \multirow{2}{*}{Poor\%} & \multirow{2}{*}{Overdose\%}  \tabularnewline
\cline{2-8} 
 &  & 1 & 2 & 3 & 4 & 5 & 6 & (in months) &  &  &   \tabularnewline
\hline 
%Scenario 1 & Pr(tox) & \textbf{0.28} & 0.42 & 0.49 & 0.61 & 0.76 & 0.87 &  &  &  &   \tabularnewline
%\cline{2-8} 
%R6 & Sel\% & \textbf{36.2} & 8.8 & 1.1 & 0.0 & 0.0 & 0.0 & 23.3 & 53.9 & 70.8 & 9.9 \tabularnewline
% & Pts\% & \textbf{40.8} & 11.9 & 2.0 & 0.2 & 0.0 & 0.0 &  &  &  &   \tabularnewline
%TITE-CRM & Sel\% & \textbf{54.0} & 25.2 & 2.6 & 0.1 & 0.0 & 0.0 & 20.5 & 18.1 & 10.4 & 28.3 \tabularnewline
% & Pts\% & \textbf{56.8} & 25.7 & 6.2 & 1.0 & 0.1 & 0.0 &  &  &  &   \tabularnewline
%TITE-mTPI & Sel\% & \textbf{67.8} & 21.8 & 3.3 & 0.2 & 0.0 & 0.0 & 22.2 & 6.9 & 13.2 & 30.4 \tabularnewline
% & Pts\% & \textbf{65.0} & 25.7 & 5.1 & 0.5 & 0.0 & 0.0 &  &  &  &   \tabularnewline
%TITE-keyboard & Sel\% & \textbf{61.1} & 23.9 & 3.6 & 0.3 & 0.0 & 0.0 & 22.9 & 11.1 & 5.3 & 25.0 \tabularnewline
% & Pts\% & \textbf{61.4} & 25.7 & 6.0 & 1.0 & 0.1 & 0.0 &  &  &  &   \tabularnewline
% %Keyboard & Sel\% & \textbf{60.0} & 22.3 & 3.7 & 0.2 & 0.0 & 0.0 & 44.8 & 13.8 & 9.0 & 26.5 & 10.9\tabularnewline
% %& Pts\% & \textbf{58.3} & 25.3 & 6.1 & 0.8 & 0.1 & 0.0 &  &  &  &  & \tabularnewline
%TITE-keyboard$^{{\rm t}}$ & Sel\% & \textbf{62.4} & 22.7 & 3.3 & 0.3 & 0.0 & 0.0 & 22.8 & 11.4 & 4.6 & 22.9 \tabularnewline
% & Pts\% & \textbf{63.1} & 23.7 & 5.4 & 0.8 & 0.1 & 0.0 &  &  &  &   \tabularnewline
%CRM & Sel\% & \textbf{52.5} & 25.8 & 3.0 & 0.1 & 0.0 & 0.0 & 43.6 & 18.6 & 11.0 & 29.6 \tabularnewline
% & Pts\% & \textbf{54.5} & 24.9 & 7.7 & 1.1 & 0.1 & 0.0 &  &  &  &  \tabularnewline
% &  &  &  &  &  &  &  &  &  &  &   \tabularnewline

Scenario 1 & Pr(tox) & 0.13 & \textbf{0.28} & 0.41 & 0.50 & 0.60 & 0.70 &  &  &  &   \tabularnewline
\cline{2-8} 

R6 & Sel\% & 43.2 & \textbf{30.5} & 7.8 & 0.8 & 0.0 & 0.0 & 35.9 & 17.7 & 70.8 & 10.0 \tabularnewline
 & Pts\% & 45.6 & \textbf{29.0} & 9.0 & 1.5 & 0.1 & 0.0 &  &  &  &   \tabularnewline
TITE-CRM & Sel\% & 8.4 & \textbf{58.7} & 29.2 & 2.7 & 0.1 & 0.0 & 20.5 & 0.9 & 10.4 & 28.3 \tabularnewline
 & Pts\% & 25.2 & \textbf{42.6} & 25.1 & 5.7 & 0.7 & 0.0 &  &  &  &   \tabularnewline
TITE-mTPI & Sel\% & 14.7 & \textbf{58.8} & 22.8 & 3.2 & 0.2 & 0.0 & 22.1 & 0.3 & 13.2 & 30.3 \tabularnewline
 & Pts\% & 29.4 & \textbf{46.0} & 20.0 & 3.9 & 0.4 & 0.0 &  &  &  &   \tabularnewline
TITE-keyboard & Sel\% & 13.9 & \textbf{58.2} & 23.2 & 4.0 & 0.4 & 0.0 & 22.9 & 0.3 & 5.3 & 25.0 \tabularnewline
 & Pts\% & 33.3 & \textbf{41.9} & 19.3 & 4.5 & 0.7 & 0.1 &  &  &  &   \tabularnewline
% Keyboard & Sel\% & 15.2 & \textbf{56.0} & 23.7 & 4.1 & 0.3 & 0.0 & 44.8 & 0.8 & 9.0 & 22.7 & 10.0\tabularnewline
 %& Pts\% & 27.9 & \textbf{43.6} & 21.9 & 5.2 & 0.7 & 0.0 &  &  &  &  & \tabularnewline
TITE-keyboard$^{{\rm t}}$ & Sel\% & 14.4 & \textbf{58.1} & 23.1 & 3.6 & 0.3 & 0.0 & 22.8 & 0.4 & 4.6 & 17.6 \tabularnewline
 & Pts\% & 34.2 & \textbf{42.0} & 18.7 & 4.1 & 0.6 & 0.1 &  &  &  &   \tabularnewline
CRM & Sel\% & 7.9 & \textbf{58.9} & 29.8 & 2.6 & 0.1 & 0.0 & 49.4 & 0.7 & 17.4 & 30.0 \tabularnewline
 & Pts\% & 22.9 & \textbf{42.1} & 27.2 & 6.4 & 0.8 & 0.1 &  &  &  &   \tabularnewline
 &  &  &  &  &  &  &  &  &  &  &   \tabularnewline

Scenario 2 & Pr(tox) & 0.08 & 0.15 & \textbf{0.29} & 0.43 & 0.50 & 0.57 &  &  &  &   \tabularnewline
\cline{2-8} 
R6 & Sel\% & 20.3 & 40.2 & \textbf{25.9} & 5.4 & 0.6 & 0.0 & 38.0 & 7.5 & 94.2 & 0.0 \tabularnewline
 & Pts\% & 30.3 & 35.6 & \textbf{20.6} & 6.1 & 1.0 & 0.1 &  &  &  &   \tabularnewline
TITE-CRM & Sel\% & 0.2 & 14.1 & \textbf{61.7} & 21.9 & 1.9 & 0.1 & 26.6 & 0.2 & 17.1 & 18.2 \tabularnewline
 & Pts\% & 14.0 & 23.1 & \textbf{39.5} & 19.3 & 3.6 & 0.3 &  &  &  &   \tabularnewline
TITE-mTPI & Sel\% & 1.0 & 21.8 & \textbf{56.4} & 17.7 & 2.7 & 0.3 & 26.9 & 0.0 & 26.5 & 12.7 \tabularnewline
 & Pts\% & 16.4 & 29.6 & \textbf{37.0} & 14.3 & 2.3 & 0.3 &  &  &  &   \tabularnewline
TITE-keyboard & Sel\% & 1.1 & 20.8 & \textbf{55.5} & 19.9 & 3.3 & 0.4 & 27.2 & 0.0 & 15.4 & 7.5 \tabularnewline
 & Pts\% & 17.8 & 31.5 & \textbf{33.3} & 13.8 & 3.0 & 0.4 &  &  &  &  \tabularnewline
% Keyboard & Sel\% & 1.2 & 19.4 & \textbf{55.7} & 19.6 & 3.7 & 0.4 & 50.0 & 0.1 & 15.0 & 13.7 & 9.2\tabularnewline
% & Pts\% & 13.2 & 27.4 & \textbf{38.1} & 17.1 & 3.6 & 0.6 &  &  &  &  & \tabularnewline
TITE-keyboard$^{{\rm t}}$ & Sel\% & 1.0 & 21.1 & \textbf{55.9} & 18.5 & 3.1 & 0.4 & 27.2 & 0.0 & 15.8 & 6.7 \tabularnewline
 & Pts\% & 18.1 & 32.1 & \textbf{33.2} & 13.5 & 2.7 & 0.4 &  &  &  &   \tabularnewline
CRM & Sel\% & 0.1 & 13.1 & \textbf{62.8} & 22.4 & 2.1 & 0.1 & 49.8 & 0.0 & 16.2 & 19.4 \tabularnewline
 & Pts\% & 12.3 & 20.8 & \textbf{42.5} & 20.6 & 3.8 & 0.5 &  &  &  &   \tabularnewline
 &  &  &  &  &  &  &  &  &  &  &   \tabularnewline
 
 Scenario 3 & Pr(tox) & \textbf{0.28} & 0.42 & 0.49 & 0.61 & 0.76 & 0.87 &  &  &  &   \tabularnewline
\cline{2-8} 
R6 & Sel\% & \textbf{36.2} & 8.8 & 1.1 & 0.0 & 0.0 & 0.0 & 23.3 & 53.9 & 70.8 & 9.9 \tabularnewline
 & Pts\% & \textbf{40.8} & 11.9 & 2.0 & 0.2 & 0.0 & 0.0 &  &  &  &   \tabularnewline
TITE-CRM & Sel\% & \textbf{54.0} & 25.2 & 2.6 & 0.1 & 0.0 & 0.0 & 20.5 & 18.1 & 10.4 & 28.3 \tabularnewline
 & Pts\% & \textbf{56.8} & 25.7 & 6.2 & 1.0 & 0.1 & 0.0 &  &  &  &   \tabularnewline
TITE-mTPI & Sel\% & \textbf{67.8} & 21.8 & 3.3 & 0.2 & 0.0 & 0.0 & 22.2 & 6.9 & 13.2 & 30.4 \tabularnewline
 & Pts\% & \textbf{65.0} & 25.7 & 5.1 & 0.5 & 0.0 & 0.0 &  &  &  &   \tabularnewline
TITE-keyboard & Sel\% & \textbf{61.1} & 23.9 & 3.6 & 0.3 & 0.0 & 0.0 & 22.9 & 11.1 & 5.3 & 25.0 \tabularnewline
 & Pts\% & \textbf{61.4} & 25.7 & 6.0 & 1.0 & 0.1 & 0.0 &  &  &  &   \tabularnewline
 %Keyboard & Sel\% & \textbf{60.0} & 22.3 & 3.7 & 0.2 & 0.0 & 0.0 & 44.8 & 13.8 & 9.0 & 26.5 & 10.9\tabularnewline
 %& Pts\% & \textbf{58.3} & 25.3 & 6.1 & 0.8 & 0.1 & 0.0 &  &  &  &  & \tabularnewline
TITE-keyboard$^{{\rm t}}$ & Sel\% & \textbf{62.4} & 22.7 & 3.3 & 0.3 & 0.0 & 0.0 & 22.8 & 11.4 & 4.6 & 22.9 \tabularnewline
 & Pts\% & \textbf{63.1} & 23.7 & 5.4 & 0.8 & 0.1 & 0.0 &  &  &  &   \tabularnewline
CRM & Sel\% & \textbf{52.5} & 25.8 & 3.0 & 0.1 & 0.0 & 0.0 & 43.6 & 18.6 & 11.0 & 29.6 \tabularnewline
 & Pts\% & \textbf{54.5} & 24.9 & 7.7 & 1.1 & 0.1 & 0.0 &  &  &  &  \tabularnewline
\hline 
\end{tabular}
}
\end{center}
\vspace{0.5in}
{\footnotesize Note: R6 is the rolling six design; TITE-CRM is the time-to-event CRM; TITE-keyboard and TITE-mTPI are the proposed time-to-event versions of keyboard and mTPI designs, respectively.  TITE-keyboard$^{{\rm t}}$ is the design that utilizes the true likelihood function (\ref{likelihood}).
CRM is the continual reassessment method based on the complete data.
 ``Stop\%'' is the early stopping percentage; ``Poor\%'' is the risk of poor allocation; ``Overdose\%'' is the risk of overdosing. }
\end{table}

\begin{table}
\begin{center}
{Table \ref{fixres} continued. }
 \tabcolsep 0.05in
 \renewcommand\arraystretch{1}
 {\footnotesize
\begin{tabular}{llcccccccccc}
\hline 
\multirow{2}{*}{Methods} &  & \multicolumn{6}{c}{Dose level} & Duration & \multirow{2}{*}{Stop\%} & \multirow{2}{*}{Poor\%} & \multirow{2}{*}{Overdose\%}  \tabularnewline
\cline{2-8} 
 &  & 1 & 2 & 3 & 4 & 5 & 6 & (in months) &  &  &   \tabularnewline
\hline 
Scenario 4 & Pr(tox) & 0.05 & 0.10 & 0.20 & \textbf{0.31} & 0.50 & 0.70 &  &  &  &   \tabularnewline
\cline{2-8} 

R6 & Sel\% & 10.9 & 29.9 & 33.8 & \textbf{19.8} & 2.5 & 0.0 & 38.3 & 3.2 & 99.7 & 0.0 \tabularnewline
 & Pts\% & 24.0 & 31.2 & 25.6 & \textbf{12.6} & 3.6 & 0.4 &  &  &  &  \tabularnewline
TITE-CRM & Sel\% & 0.0 & 1.5 & 29.5 & \textbf{58.9} & 9.9 & 0.1 & 28.3 & 0.1 & 23.7 & 5.7 \tabularnewline
 & Pts\% & 11.3 & 14.1 & 29.1 & \textbf{33.3} & 11.2 & 0.9 &  &  &  &   \tabularnewline
TITE-mTPI & Sel\% & 0.2 & 5.2 & 35.3 & \textbf{49.5} & 9.7 & 0.3 & 28.5 & 0.0 & 38.9 & 3.4 \tabularnewline
 & Pts\% & 12.7 & 19.7 & 31.5 & \textbf{27.1} & 8.3 & 0.7 &  &  &  &   \tabularnewline
TITE-keyboard & Sel\% & 0.2 & 4.3 & 33.2 & \textbf{49.8} & 12.0 & 0.4 & 28.8 & 0.0 & 28.1 & 1.7 \tabularnewline
 & Pts\% & 13.5 & 21.2 & 30.4 & \textbf{25.0} & 8.9 & 1.1 &  &  &  &   \tabularnewline
 %Keyboard & Sel\% & 0.2 & 4.0 & 30.9 & \textbf{52.9} & 11.6 & 0.4 & 50.2 & 0.0 & 21.3 & 2.7 \tabularnewline
 %& Pts\% & 10.4 & 16.3 & 29.8 & \textbf{31.2} & 11.1 & 1.2 &  &  &  &   \tabularnewline
TITE-keyboard$^{{\rm t}}$ & Sel\% & 0.2 & 4.5 & 35.4 & \textbf{49.1} & 10.5 & 0.3 & 28.7 & 0.0 & 30.6 & 1.3 \tabularnewline
 & Pts\% & 13.7 & 22.1 & 31.1 & \textbf{24.3} & 7.9 & 0.9 &  &  &  &   \tabularnewline
CRM & Sel\% & 0.1 & 1.4 & 29.7 & \textbf{58.9} & 9.9 & 0.1 & 50.0 & 0.0 & 20.7 & 5.5 \tabularnewline
 & Pts\% & 10.3 & 11.8 & 28.7 & \textbf{36.9} & 11.3 & 1.0 &  &  &  &   \tabularnewline
 &  &  &  &  &  &  &  &  &  &  &   \tabularnewline
Scenario 5 & Pr(tox) & 0.06 & 0.08 & 0.12 & 0.18 & \textbf{0.30} & 0.41 &  &  &  &   \tabularnewline
\cline{2-8} 

R6 & Sel\% & 7.6 & 13.6 & 21.9 & 30.7 & \textbf{17.1} & 4.8 & 36.4 & 4.4 & 100.0 & 0.0 \tabularnewline
 & Pts\% & 21.8 & 22.7 & 22.0 & 17.6 & \textbf{8.7} & 3.6 &  &  &  &   \tabularnewline
TITE-CRM & Sel\% & 0.0 & 0.4 & 5.7 & 32.2 & \textbf{47.4} & 14.2 & 30.0 & 0.1 & 42.1 & 6.5 \tabularnewline
 & Pts\% & 11.8 & 12.2 & 17.3 & 25.4 & \textbf{23.7} & 9.6 &  &  &  &  \tabularnewline
TITE-mTPI & Sel\% & 0.2 & 1.5 & 8.9 & 32.9 & \textbf{40.7} & 15.9 & 30.6 & 0.0 & 50.4 & 3.5 \tabularnewline
 & Pts\% & 13.2 & 15.5 & 20.1 & 23.9 & \textbf{19.2} & 8.2 &  &  &  &   \tabularnewline
TITE-keyboard & Sel\% & 0.1 & 0.8 & 7.5 & 30.3 & \textbf{43.3} & 18.0 & 31.0 & 0.0 & 37.4 & 0.9 \tabularnewline
 & Pts\% & 13.4 & 15.6 & 19.8 & 23.8 & \textbf{18.7} & 8.6 &  &  &  &   \tabularnewline
% Keyboard & Sel\% & 0.1 & 0.8 & 4.6 & 26.4 & \textbf{44.4} & 23.8 & 50.5 & 0.0 & 28.7 & 2.2 \tabularnewline
% & Pts\% & 10.5 & 12.2 & 16.0 & 23.8 & \textbf{24.0} & 13.6 &  &  &  &   \tabularnewline
TITE-keyboard$^{{\rm t}}$ & Sel\% & 0.1 & 1.3 & 7.9 & 32.5 & \textbf{42.0} & 16.2 & 30.8 & 0.0 & 41.3 & 0.9 \tabularnewline
 & Pts\% & 13.7 & 16.2 & 20.6 & 24.2 & \textbf{17.9} & 7.5 &  &  &  &   \tabularnewline
CRM & Sel\% & 0.0 & 0.4 & 4.6 & 30.7 & \textbf{49.4} & 15.0 & 50.4 & 0.0 & 37.0 & 7.3 \tabularnewline
 & Pts\% & 10.5 & 9.9 & 15.0 & 27.2 & \textbf{26.5} & 10.8 &  &  &  &   \tabularnewline
 &  &  &  &  &  &  &  &  &  &  &   \tabularnewline
Scenario 6 & Pr(tox) & 0.05 & 0.06 & 0.08 & 0.11 & 0.19 & \textbf{0.32} &  &  &  &   \tabularnewline
\cline{2-8} 
R6 & Sel\% & 4.3 & 7.1 & 12.1 & 23.7 & 31.4 & \textbf{18.4} & 36.0 & 3.1 & 100.0 & 0.0  \tabularnewline
 & Pts\% & 19.6 & 19.7 & 19.5 & 18.2 & 12.3 & \textbf{8.2} &  &  &  &   \tabularnewline
TITE-CRM & Sel\% & 0.0 & 0.1 & 2.1 & 10.9 & 38.0 & \textbf{48.8} & 32.0 & 0.1 & 49.7 & 0.0 \tabularnewline
 & Pts\% & 11.2 & 11.0 & 13.7 & 17.8 & 23.5 & \textbf{22.9} &  &  &  &  \tabularnewline
TITE-mTPI & Sel\% & 0.1 & 0.5 & 2.6 & 11.2 & 38.0 & \textbf{47.7} & 32.6 & 0.0 & 48.9 & 0.0 \tabularnewline
 & Pts\% & 12.1 & 13.2 & 15.2 & 17.6 & 21.8 & \textbf{20.1} &  &  &  &   \tabularnewline
TITE-keyboard & Sel\% & 0.1 & 0.3 & 1.7 & 9.9 & 38.5 & \textbf{49.5} & 32.8 & 0.0 & 45.0 & 0.0 \tabularnewline
 & Pts\% & 12.2 & 13.1 & 15.2 & 18.8 & 21.7 & \textbf{18.9} &  &  &  &   \tabularnewline
% Keyboard & Sel\% & 0.0 & 0.2 & 0.7 & 5.8 & 33.4 & \textbf{60.0} & 50.7 & 0.0 & 25.6 & 0.0 \tabularnewline
% & Pts\% & 10.0 & 10.7 & 11.9 & 15.2 & 23.5 & \textbf{28.7} &  &  &  &   \tabularnewline
TITE-keyboard$^{{\rm t}}$ & Sel\% & 0.0 & 0.2 & 1.8 & 10.4 & 38.9 & \textbf{48.7} & 32.7 & 0.0 & 47.4 & 0.0 \tabularnewline
 & Pts\% & 12.2 & 13.2 & 15.5 & 19.1 & 21.7 & \textbf{18.3} &  &  &  &   \tabularnewline
CRM & Sel\% & 0.0 & 0.1 & 1.1 & 8.6 & 40.7 & \textbf{49.5} & 50.6 & 0.0 & 42.7 & 0.0  \tabularnewline
 & Pts\% & 10.0 & 9.2 & 11.4 & 17.6 & 25.8 & \textbf{26.0} &  &  &  &   \tabularnewline
\hline 
\end{tabular}

}
\end{center}
\vspace{0.5in}

{\footnotesize Note: R6 is the rolling six design; TITE-CRM is the time-to-event CRM; TITE-keyboard and TITE-mTPI are the proposed time-to-event versions of keyboard and mTPI designs, respectively.  TITE-keyboard$^{{\rm t}}$ is the design that utilizes the true likelihood function (\ref{likelihood}).
CRM is the continual reassessment method based on the complete data.
 ``Stop\%'' is the early stopping percentage; ``Poor\%'' is the risk of poor allocation; ``Overdose\%'' is the risk of overdosing. }
\end{table}

\end{document}